\documentclass[amssymb,prl,aps,floatfix,floats,preprintnumbers,showpacs,showkeys,twocolumn,superscriptaddress,groupedaddress]{revtex4-1}
\usepackage{dcolumn}
\usepackage{bm}
\usepackage{amssymb}
\usepackage{rcs}
\usepackage{color}
\usepackage{graphicx}
\usepackage{graphics}
\usepackage[normalem]{ulem}
\usepackage{chngcntr}
\usepackage[thinspace,thinqspace]{SIunits}
\usepackage{natbib}
\usepackage{amsmath}
\newcommand{\sro}{Sr$_{3}$Ru$_{2}$O$_{7}$}
\begin{document}
\preprint{APS/123-QED}
\title{Low temperature thermodynamic investigation of the phase diagram of \sro}
\author{D.~Sun$^1$, A.~W.~Rost$^{2,3,4}$, R.~S.~Perry$^5$, A.~P.~Mackenzie$^{1,4}$ and M.~Brando$^{1}$}
\affiliation{
	$^{1}$Max Planck Institute for Chemical Physics of Solids, D-01187 Dresden, Germany\\
	$^{2}$ Max Planck Institute for Solid State Research, D-70569 Stuttgart, Germany\\
    $^{3}$ Institute for Functional Matter and Quantum Technologies, University of Stuttgart, Pfaffenwaldring 57, D-70569 Stuttgart, Germany\\
	$^{4}$ Scottish Universities Physics Alliance, School of Physics and Astronomy, University of St.\ Andrews, St.\ Andrews, Fife KY16 9SS, UK\\
	$^{5}$ London Centre for Nanotechnology and Department of Physics and Astronomy, University College London, London WC1E 6BT, UK
}
\date{\today}
%\footnote{Andy.Mackenzie@cpfs.mpg.de, Manuel.Brando@cpfs.mpg.de}
%
\begin{abstract}
We studied the phase diagram of \sro\ by means of heat capacity and magnetocaloric effect measurements at temperatures as low as 0.06\,K and fields up to 12\,T. We confirm the presence of a new quantum critical point at 7.5\,T which is characterized by a strong non-Fermi-liquid behavior of the electronic specific heat coefficient $\Delta C/T \sim -\log T$ over more than a decade in temperature, placing strong constraints on theories of its criticality. In particular logarithmic corrections are found when the dimension $d$ is equal to the dynamic critical exponent $z$, in contrast to the conclusion of a two-dimensional metamagnetic quantum critical endpoint, recently proposed. Moreover, we achieved a clear determination of the new second thermodynamic phase adjoining the first one at lower temperatures. Its thermodynamic features differ significantly from those of the dominant phase and characteristics expected of classical equilibrium phase transitions are not observed, indicating fundamental differences in the phase formation.
\end{abstract}
\maketitle
\section{Introduction} 
\begin{figure}[t]
	\centering
	\includegraphics[width=\columnwidth]{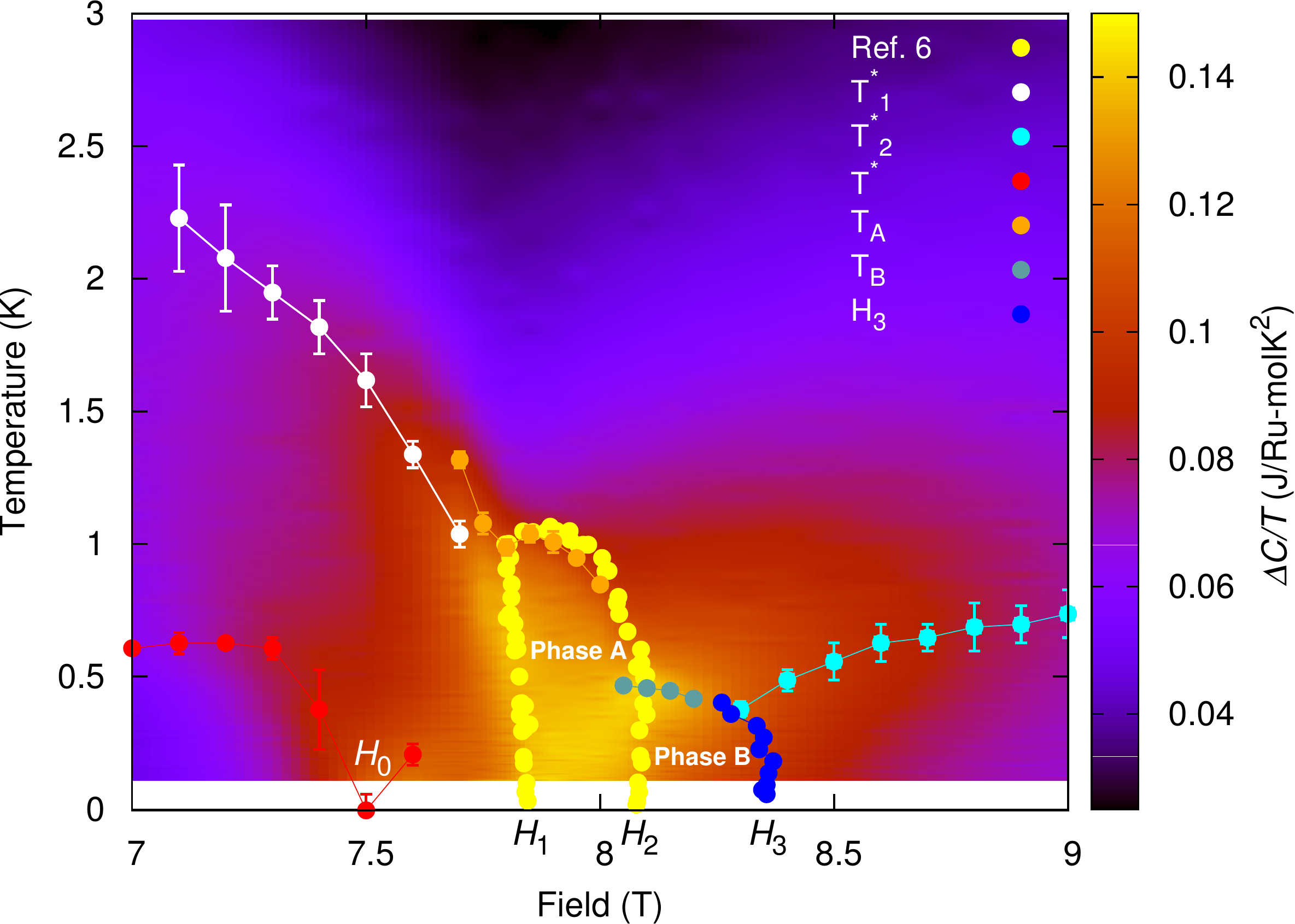}
	\caption{Phase diagram of \sro\ with $H \parallel c$ showing the locations in the ($T,H$) plane of the key thermodynamic features deduced from the specific heat and magnetocaloric measurements described here and those published in Ref.~\cite{grigera2004disorder}. In the background we show a color plot of interpolated $\Delta C/T = (C­(T,H)-C(T,0))/T$ data to show the temperature and field development of the energy scales indicated by the points. $T^{*}_{1}$ is a crossover identified from a maximum in $\Delta C/T$ (Fig.~\ref{fig3}a for $7.3-7.6$\,T). $T^*_2$ is a similarly defined crossover deduced from $\Delta C/T$ data for $8.3-9$\,T.  $T^*$ is a second crossover scale as indicated in Fig.~\ref{fig3}a. $T_{A}$ and $T_{B}$ are identified from features in $\Delta C/T$ as labeled in Fig.~\ref{fig4}a,b and d. $H_{3}$ is identified from magnetocaloric data as shown in Fig.~\ref{fig4}c. See also Fig.~\ref{fig3_appendix} of the Appendix.}
	%The extent to which the final three features can be interpreted as marking phase boundaries is discussed in the main text.
	\label{fig1}       
\end{figure}

The formation of new phases and the emergence of quantum critical points (QCPs) play a key role in the phase diagrams of a wide range of strongly correlated electron systems. Since first being synthesized in single crystal form \cite{cao1997observation,ikeda2000ground}, the layered perovskite metal \sro\ has been the subject of intense study due to its peculiar correlated electron properties \cite{mackenzie2012quantum}. In particular an unusual phase stabilized in the vicinity of an underlying metamagnetic quantum critical end point has attracted significant interest \cite{mackenzie2012quantum}. The fine balance of the energetics involved in the formation of this phase due to competing interactions at the quantum critical point in \sro\ is evidenced by the dependence of the observed properties on sample purity. Single crystals can be grown in image furnaces, and, with care, residual resistivities $\rho_0$ below 1 $\mu\Omega$cm can be achieved \cite{perry2004systematic}. For $\rho_0$ in the range 3-5\,$\mu\Omega$cm, a single metamagnetic transition is observed. For fields applied in the \textit{­ab} plane the transition occurs near 5\,T with a magnitude of $\sim$ 0.25\,$\mu_{\textrm{B}}$/Ru and is first-order below a critical end-point at approximately 1.2\,K.  As the field is rotated to the \textit{c} axis, the metamagnetic field rises to nearly 8\,T and the end-point temperature falls to below 100 mK \cite{grigera2003angular}.  If $\rho_0$ is reduced to below 1 $\mu\Omega$cm, however, qualitatively different behavior is seen, with the phase diagram as currently known being summarized in Fig.~\ref{fig1}. Most prominently, a new phase (labelled `A')  bounded in field by first-order phase transitions at 7.8 and 8.1\,T and in temperature by a second-order transition at 1.1\,K was reported \cite{grigera2004disorder,rost2009entropy,gegenwart2006metamagnetic}, with signatures of another metamagnetic feature at 7.5\,T.

More recent work on the latest generation of samples revealed evidence for a putative second phase (`B') extending from 8.1 to 8.5 T, with a lower onset \textit{T}$_c$ of less than 0.6 K \cite{stingl2013electronic,bruin2013study}.  This was shown particularly clearly in breakthrough neutron scattering measurements that established incommensurate order with $Q = (0.233,0,0)$ within the A phase and $Q  = (0.218,0,0)$ within the B phase. In both cases the correlation length of the ordered signal was greater than 350\,\AA\  and the characteristic frequency of any fluctuations less than 1 GHz \cite{lester2015field}.

This phase diagram is unusual in several ways. Firstly, the application of a uniform magnetic field stimulates the formation of phases that feature finite $q$ order. Secondly, the resistivity rises substantially over the background value in both enclosed phases. Thirdly, in-plane transport in these phases has a giant susceptibility to anisotropy \cite{bruin2013study,borzi2007formation} that can be stimulated both by in-plane magnetic fields and by in-plane uniaxial strain \cite{brodsky2017strain}.  Finally, the curvature of the first-order transition lines just above $H_{1}$ and $H_{2}$ implies that the entropy within the A phase is higher than that at lower or higher fields.  Although not unprecedented (for example such phenomenology is at the root of the Pomeranchuk effect in $^3$He) this is unexpected. 

The unusual nature of the phase diagram motivated detailed studies of the electronic properties of this material by, e.g., band structure calculations~\cite{hase1997electronic,singh2001electronic}, angle-resolved photoemission spectroscopy (ARPES)~\cite{tamai2008fermi,allan2013formation}, de Haas-van Alphen (dHvA)~\cite{borzi2004haas,mercure2010quantum} and magnetic Gr\"uneisen parameter measurements~\cite{tokiwa2016multiple}.
%
%The unusual nature of the phase diagram motivated detailed studies of the electronic properties of this material. As the $n = 2$ member of the Ruddlesden-Popper series Sr$_{n+1}$Ru$_n$O$_{3n+1}$, conduction occurs in RuO$_2$ planes stacked as bilayers.  Band structure calculations show that because of a symmetry-lowering rotation of Ru-O octahedra, the bands crossing the Fermi level have character deriving from four different Ru $4­d$ orbitals: $d_{xy}$, $d_{xz}$, $d_{yz}$ and $d_{x^2-y^2}$\cite{hase1997electronic,singh2001electronic}. These calculations, as well as angle-resolved photoemission spectroscopy (ARPES) and de Haas-van Alphen (dHvA) measurements \cite{tamai2008fermi,allan2013formation,borzi2004haas,mercure2010quantum}, show that the bilayer coupling is reasonably strong, leading to a pronounced bilayer splitting of the parts of the Fermi surface with strong $d_{xz}$, $d_{yz}$ character.  In contrast, inter-bilayer coupling is weak, and electrical transport is strongly anisotropic, with a factor of approximately 300 between the inter- and in-plane resistivity \cite{ikeda2000ground}.
%
Although the band structure calculations correctly predict the basic topography of the Fermi surface, the ARPES experiments have shown that \sro\ is strongly renormalized, with bandwidths a factor of $7-20$ narrower than the calculated values~\cite{tamai2008fermi,allan2013formation}.  This is reflected in the specific heat, which is 0.11\,J/Ru-molK $^2$ in zero applied magnetic field~\cite{ikeda2000ground}. The strong correlations implied by these observations are also evident in the magnetic properties.  The $q = 0$ magnetic susceptibility is large, corresponding to a Wilson ratio of 10 and suggesting that \sro\ is on the border of ferromagnetism.  Indeed, modest uniaxial pressure can drive it ferromagnetic~\cite{ikeda2004uniaxial}, and applied fields in the range $5-8$\,T lead to metamagnetism~\cite{perry2001metamagnetism}. In contrast to uniaxial pressure, hydrostatic pressure weakens the magnetism, as would be qualitatively expected in a Stoner picture in which sharp features in the density of states near the Fermi level are at the root of the strong correlations and magnetism~\cite{chiao2002effect,wu2011quantum,sun2013pressure}. However, the situation seems to be more complex: In a work carried out in parallel to this work, Tokiwa \textit{et al.} have established the existence of a second metamagnetic quantum critical end-point (QCEP) at about 7.5\,T ($H_{0}$ in Fig.~\ref{fig1}) in addition to the previously known one near 7.85\,T and proposed quantum critical regimes of both instabilities. According to this analysis and the field dependence of the Sommerfeld coefficient, the authors suggested that the nature of the QCP near 7.85\,T is that of a two-dimensional (2D) QCEP~\cite{tokiwa2016multiple,millis2002metamagnetic}.

Overall these findings establish \sro\ to be a quasi-2D strongly correlated metal in which magnetic interactions play a crucial role. The role and interplay of the two QCEPs in controlling the low temperature / high field phase diagram and in particular the question of the nature of the phases stabilized in the quantum critical regime clearly merit further detailed studies. Although a first generation of thermodynamic experiments established the phase boundaries of the A phase, it revealed no strong signatures of the B phase~\cite{rost2009entropy}.  In addition, the experiments were cut off for technical reasons below 0.2\,K, an uncomfortably high temperature when studying a phase diagram with a characteristic temperature scale of $0.5-1$\,K. The purpose of the current work is to study the specific heat and magnetocaloric effect in \sro\ in more detail, at higher resolution and at lower temperatures than in the previous work.  We show that thermodynamic signatures of the B phase can be resolved, but that they are much weaker than those of the A phase indicating significant differences in the order parameter. Crucially a detailed quantitative analysis shows that they cannot unambiguously be associated with classical equilibrium phase transitions raising the question of the role of (quantum-) fluctuations.  Our low temperature measurements also reveal that a thermodynamic feature at 7.5\,T (previously assumed to be a crossover because of its width in field of 0.2\,K) in fact has the characteristics of a zero temperature QCEP related to a lower energy scale than those previously identified as being crucial to the physics of \sro. Intriguingly the temperature evolution of this new quantum critical regime is cut off at a scale associated with the dominant critical point at 8\,T, implying a clear hierarchy of the energy scales and thereby making \sro\ a rare example of a system with multiple quantum phase transitions. The nature of the new QCEP is discussed in light of the recent observation of quantum critical scaling in the magnetic Gr\"uneisen parameter~\cite{tokiwa2016multiple}. %
\section{Results}
Single crystals used in the measurements reported here were grown and characterized using the methods described in Refs.~\cite{perry2004systematic,ikeda2002bulk} and have a residual resistivity $\rho_0$ = 0.5\,$\mu\Omega$cm. Specific heat was measured using the compensated heat pulse method~\cite{Wilhelm2004}. Experiments were performed at 37 different fixed fields spanning 0 to 12\,T, while varying temperature from 0.06 to 4\,K. The data shown in the paper have had Schottky, phonon and addenda contributions subtracted. Magnetocaloric measurements were performed in two different experimental setups, optimized to study the magnetocaloric effect in opposite limits.  The first was an adaptation of the calibrated non-adabiatic technique employed in Ref.~\cite{rost2009entropy}, but in new apparatus with a reduced base temperature of approximately 0.12\,K. The second was quasi-adiabatic, with the advantage of enabling work at lower temperatures down to approximately 0.06\,K.  Field sweep rates were 5 and 10\,mT/min for the respective experiments.
 
\begin{figure}[b]
\begin{center}
\includegraphics[width=\columnwidth]{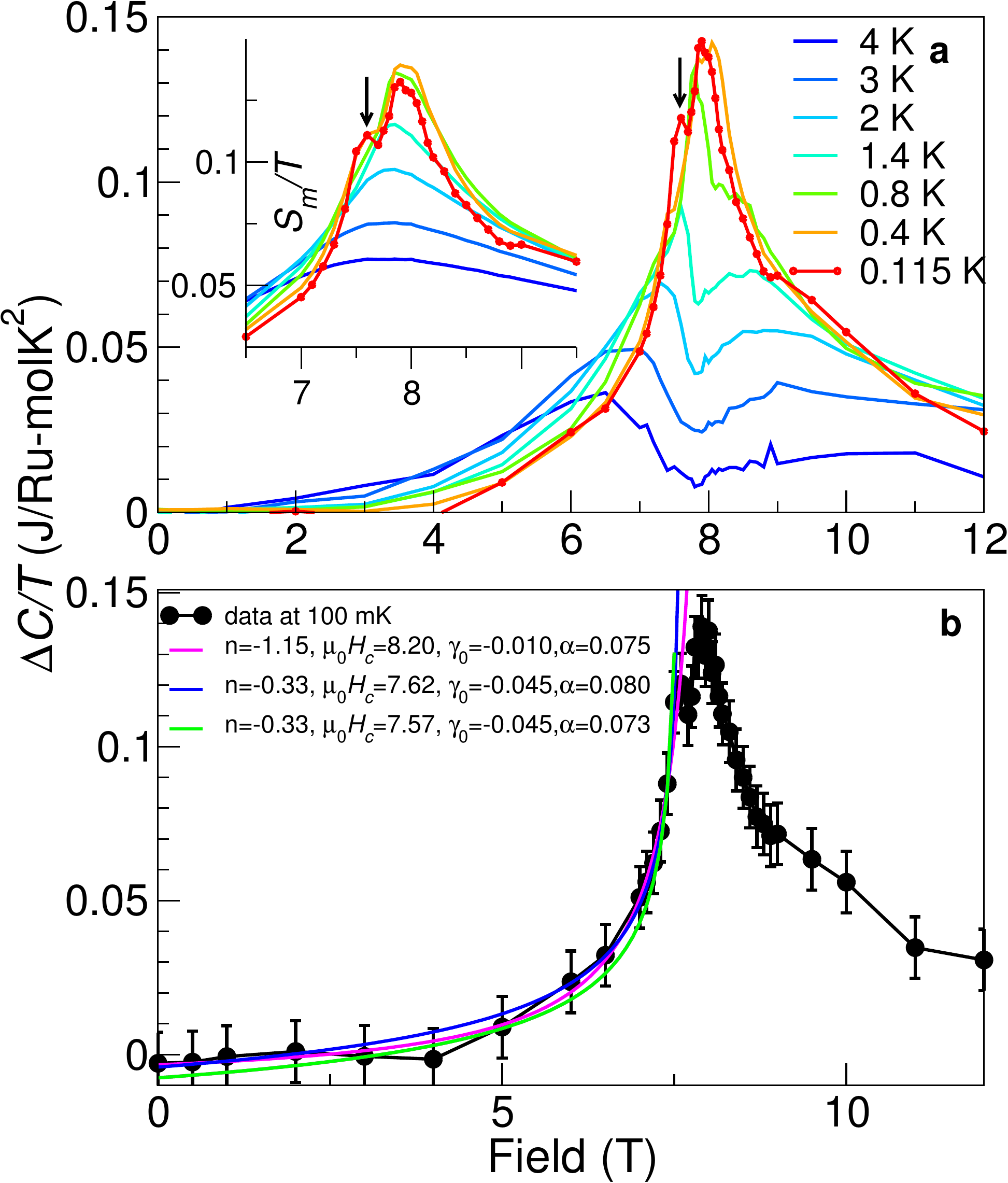}
\end{center}
\caption{(a) Electronic specific heat $\Delta C/T$ of \sro\ as a function of magnetic field for select temperatures. Curves are generated from $T$-sweeps at constant field.  Magnetic entropy $S_{m}/T$ as a function of magnetic field calculated from these data is shown in the inset. (b) Fits of the electronic specific heat coefficient $\Delta\gamma = \Delta C/T$ at 100\,mK with a function of the form $\Delta\gamma(H) = \gamma_{0} + \alpha(\mu_{0}(H_{c}-H))^{n}$. Fit 1 was performed with free parameters within the field range between 0 and 7.4\,T as done in Ref.~\cite{rost2009entropy}. The fit yields $n = -1.15$, $\mu_{0}H_{c} = 8.20$\,T, $\gamma_{0} = -0.01$\,J/Ru-molK$^2$ and $\alpha = 0.075$ (magenta line).  Fit 2 was performed in the same range but with fixed exponent $n = -1/3$ as done in Ref.~\cite{tokiwa2016multiple}: This fit yields $n = -1/3$, $\mu_{0}H_{c} = 7.62$\,T, $\gamma_{0} = -0.045$\,J/Ru-molK$^2$ and $\alpha = 0.08$ (blue line). The green curve is the fit in Ref.~\cite{tokiwa2016multiple}.}
\label{fig2}
 \end{figure}

In Fig.~\ref{fig2} we summarize our results for the field dependence of the electronic specific heat coefficient $\gamma(H) = C(T,H)/T$ of \sro. In particular, we show in Fig.~\ref{fig2}a $\Delta C/T = (C­(T,H)-C(T,0))/T$ with $\gamma_{0} = \lim_{T \to 0} C(T,0)/T = 0.103 \pm 0.005$\,J/Ru-molK$^2$ for magnetic fields between 0 and 12\,T, at a series of constant temperatures from 4 to 0.1\,K. The evolution makes an interesting comparison with the detailed data for the temperature dependence of $C/T$ presented in Refs.~\cite{rost2009entropy} and \cite{tokiwa2016multiple}. All data sets show the consistent picture that the degrees of freedom from which the unusual low temperature states form exist at high temperatures and low fields, and `pile up' at low temperature and fields around 8\,T, suggesting an association between phase formation and quantum criticality centered on approximately that field. The field evolution and sharpening of the broad peak seen at 4\,K is the signature of that process in these field-dependent measurements of $\Delta C/T$.
 
Calculation of the entropy $S$ is a particularly informative way of highlighting the underlying physics in systems like \sro. Since we took temperature dependent data at a large number of closely spaced fixed fields we have in principle the opportunity to determine the entropy as a function of field and temperature. However, integration from $T = 0$ involves assumptions about the behavior below the lowest measurement temperature. If the metallic state is a known Fermi liquid, this is a safe procedure as long as the measurements extend to sufficiently low temperatures that $\gamma$ has become temperature-independent, allowing for a trivial extrapolation of $C/T$ to $T=0$. Quantum criticality, in contrast, is often associated with a logarithmic divergence of $C/T$, invalidating the use of the Fermi liquid assumption in calculating $S$. Previous work on \sro\ employed a combination of specific heat and fully calibrated measurements of the magnetocaloric effect to establish a Fermi liquid specific heat $-$ entropy relationship below 0.25\,K for  $0 < \mu_{0}H < 7.3$\,T and $\mu_{0}H > 8.5$\,T~\cite{,rost2009thesis,rost2009entropy}. In the inset of Fig.~\ref{fig2}a we show the magnetic entropy $S_{m}/T$ calculated from our specific heat data using a Fermi liquid assumption for the extrapolation of $\Delta C/T$ below 0.1\,K. The values shown can be expected to be correct for the field ranges mentioned above, and to slightly underestimate $S_{m}$ for $7.3 < \mu_{0}H < 8.5$\,T. They illustrate the important point (confirming that reported in Ref.~\cite{rost2009entropy}) that at all measured temperatures, the entropy peak centres at approximately 8\,T, and becomes sharper as the temperature is decreased.

\begin{figure}[b]
	\begin{center}
		\includegraphics[width=\columnwidth]{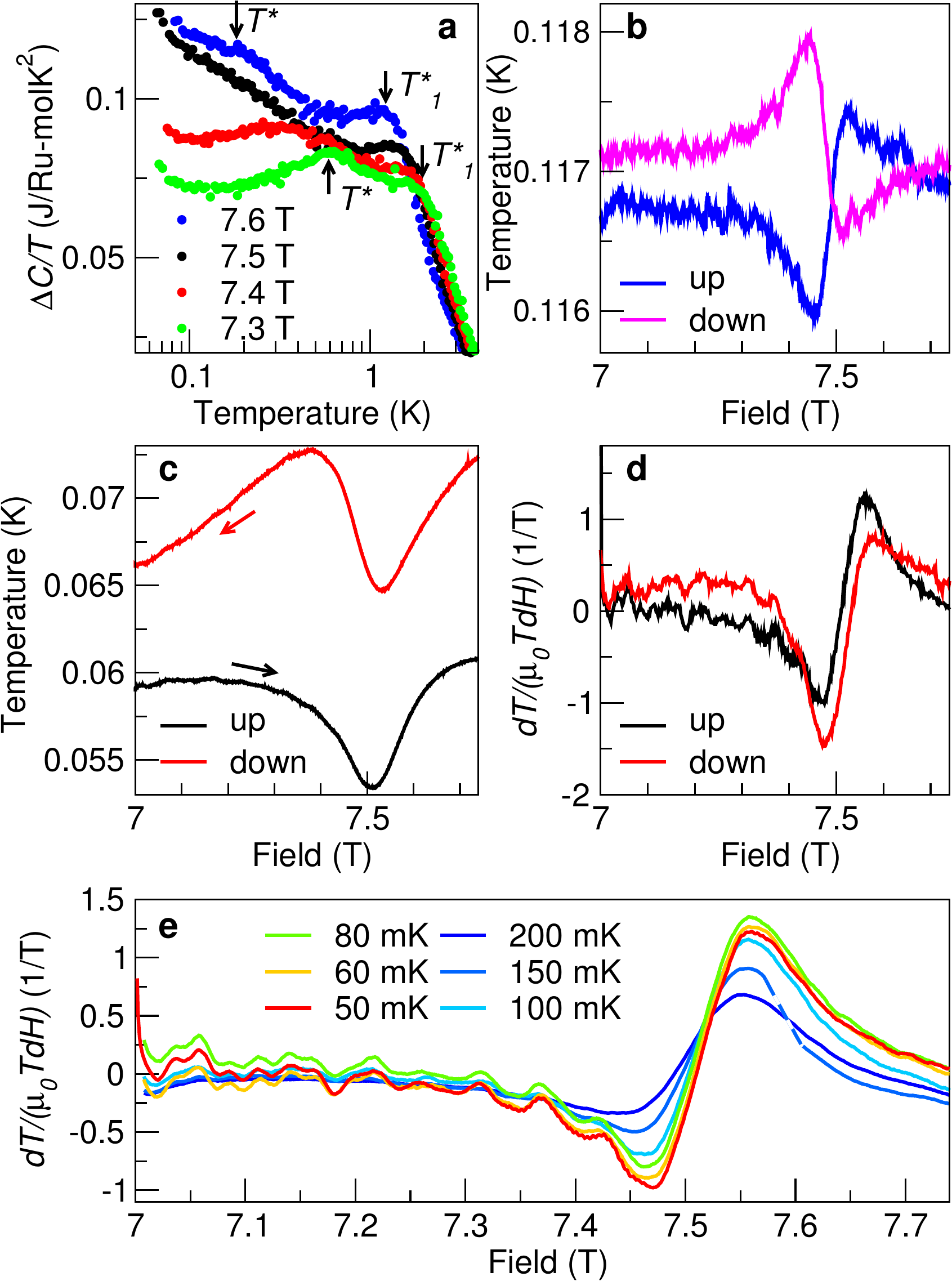}
	\end{center}
	\caption{(a) $T$-dependence of $\Delta C/T$ for fields in the region of $\mu_{0}H_{0} = 7.5$\,T.  At $H_{0}$, $\Delta C/T$ diverges to the lowest temperature of measurement, and the states involved in this divergence are seen from scans at 7.3, 7.4 and 7.6 T to be related to the depression of an energy scale $k_{B}T^{*}$ towards $T = 0$.  Magnetocaloric data using non-adiabatic and quasi-adiabatic methods are shown in (b) and (c) for $7 \leq \mu_{0}H \leq 7.8$\,T. The hysteresis seen in panel (c) is caused by irreversible heat introduced at higher fields in the same data-taking run. This difference in thermal  history between the two sweeps is related to the first order phase transitions at $H_{1}$ and $H_{2}$ and also to possible friction between the sample and its mounting plate and glue, due to large magnetostriction in the region $7.8-8.1$\,T. (d) Magnetic Gr\"{u}neisen parameter obtained from the field derivative of the $T(H)$ data.
	}
	\label{fig3}       
\end{figure}

The lowest temperature data of Fig.~\ref{fig2}a show that, below 0.4\,K, a secondary sharp peak in $\Delta C/T$ (indicated by a black arrow) emerges on the low-field side of the main peak, centred on a field of about 7.5 T.  This peak had previously been identified, and associated with a metamagnetic crossover since it is linked to a rise in magnetic moment and a peak in entropy \cite{grigera2004disorder,rost2009entropy}. The previous base temperature of 0.25\,K for the entropy measurements did not, however, give the chance to distinguish between a crossover and a very low temperature continuous phase transition.  Motivated by the sharpness of the peak seen in $\Delta C/T$ at 0.1\,K, we investigated the range of fields close to 7.5\,T using both specific heat and magnetocaloric effect measurements. In Fig.~\ref{fig3}a we show the temperature evolution of the specific heat at 7.3, 7.4, 7.5 and 7.6\,T.  Above 1.5 K, the data show the same strong logarithmic divergence previously reported for fields above the A phase, but then something surprising happens.  The data show a pronounced kink at $T^{*}_{1}$ followed, at 7.3, 7.4 and 7.6\,T, by a broad peak at $T^{*}$ = 0.6, 0.4 and 0.2\,K respectively. At 7.5\,T, a second logarithmic divergence is seen from 1\,K to our lowest temperature of measurement at 0.06\,K and is characterized by a large increase in $\Delta C/T$ of about 50\,mJ/Ru-molK$^2$ (black points in Fig.~\ref{fig3}a and Fig.~\ref{fig3_appendix}d of the Appendix). This lower temperature logarithmic divergence has the same phenomenology as the behavior seen at much higher temperatures \cite{rost2011thermodynamics}: degrees of freedom appearing as field-dependent peaks in $\Delta C(T)/T$ turn into the logarithmic divergence as their characteristic temperatures $T^*$ are lowered on the approach (from both high and low fields) to the critical field of 7.5\,T. However, the energy scale involved is lower, the number of degrees of freedom smaller and the magnitude of the divergence lower than for the main phase diagram, suggestive of a separate, `second stage' quantum critical point as also observed in Ref.~\cite{tokiwa2016multiple}. This is further born out by the observation that thermodynamic signatures of this new critical point are only appearing below the characteristic $T^*$ (green points in Fig.~\ref{fig1}) which sets the relevant energy scale associated with the dominant QCEP at 8\,T.

To examine the thermodynamic properties in the region of this critical point in more depth, we carried out magnetocaloric measurements in two different limits.  First (Fig.~\ref{fig3}b) we repeated the non-adiabatic measurement introduced in Ref.~\cite{rost2009entropy} but using improved techniques that gave a base temperature of 0.117\,K.  At this temperature, the characteristic features are already quite narrow in field, but to obtain magnetocaloric information to an even lower temperature we performed quasi-adiabatic runs shown in Figs.~\ref{fig3}c,d and e. In contrast to the non-adiabatic technique, the essential features of the data across the relevant field range have the same sign, and the magnetic Gr\"{u}neisen parameter $\Gamma_{H}= (1/T)(dT/dH)_{S}$ (Fig.~\ref{fig3}d) has the functional form characteristic of proximity to a quantum critical point~\cite{tokiwa2016multiple,zhu2003universally,garst2005sign,lohneysen2007fermi}. We also note that our data show a strong temperature dependence below 0.2\,K, and that a careful analysis reveals a continuation of the sharpening up even below 0.08\,K (Fig.~\ref{fig3}e). This last data set also illustrates why, at the previous study's base temperature of 0.2\,K~\cite{rost2009thesis,rost2009entropy}, the most plausible interpretation of the Gr\"{u}neisen parameter data was in terms of a crossover related to a critical point located in phase space well below $T = 0$, as proposed for instance in CeRu$_2$Si$_2$ \cite{weickert2010universal}. In contrast, our new data and in particular the combination of specific heat and magnetocaloric measurements give good evidence for the existence of a previously overlooked QCP in \sro\ at 7.5\,T consistent with a very recent study of the magnetic Gr\"{u}neisen parameter~\cite{tokiwa2016multiple}. It is worth noting that a quantitative analysis of the Gr\"{u}neisen parameter, e.g. the determination of the prefactor $G_{r}$ which is given by a simple combination of critical exponents~\cite{garst2005sign}, is impeded by the pronounced quantum oscillations at low fields (see Fig. \ref{fig3}e) and the presence of first order phase transitions at $H_{1}$ and $H_{2}$ (cf. Fig.~\ref{fig1}).

Finally with regard to this new low field QCP we would like to note a peculiar behavior in the observed prominent quantum oscillations upon traversing the critical field (see Fig.~\ref{fig3}e). The repeatable oscillations seen between 5 and 7.45\,T (see also Fig.~\ref{fig1_appendix} of the Appendix) are the result of entropy oscillations with a main frequency of about 450\,T, one of the known quantum oscillation frequencies of \sro\ already observed in magnetocaloric effect measurements~\cite{mercure2010quantum}, and their observation in these measurements emphasizes the high quality of the sample. These oscillations are either absent or of much smaller amplitude between 7.6 and 7.8\,T. They eventually reappear at higher fields $H > H_{3}$ (see middle panel of Fig.~\ref{fig1_appendix}). Taken at face value, this might indicate a Lifshitz transition involving part of the Fermi surface, but this hypothesis needs to be checked further. 
%
%In particular, the dominant higher frequency seen in previous dHvA measurements~\cite{mercure2010quantum,mercure2008} is not strongly changing on traversing this region of fields indicating that not all Fermi surface parts are affected equally. The current work motivates a more detailed re-examination of the dHvA effect in \sro\ in this field regime. 
%
\begin{figure}[t]
	\centering
	\includegraphics[width=\columnwidth]{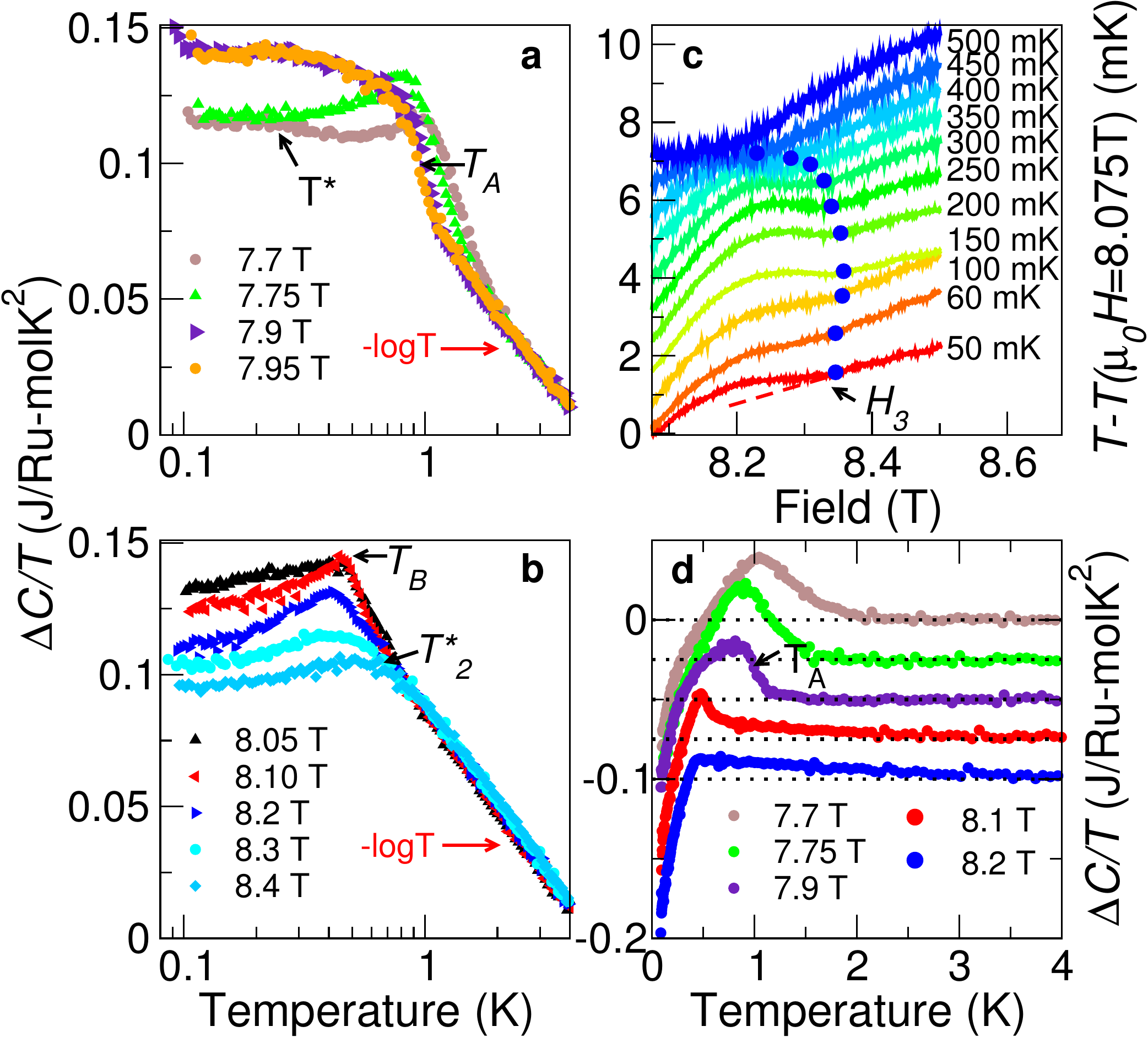}
	\caption{(a) The electronic specific heat coefficient $\Delta C/T$ as a function of temperature at 7.7, 7.75, 7.9 and 7.95\,T in the vicinity of the A phase, and (b) at 8.05, 8.1, 8.2, 8.3 and 8.4\,T, in the vicinity of the B phase. (c) Magnetocaloric data for $8.075 \leq \mu_{0}H \leq 8.5$\,T and $0.05 \leq T \leq 0.5$\,K. The blue circles mark the locations $H_{3}$ of a weak feature (kink) that we tentatively associate with the boundary of the B phase, but the data show no obvious signature of a phase transition. (d) $\Delta C/T$ across the region of phase 	formation after subtraction of the logarithmically diverging background signal extrapolated to low temperatures, following the procedure described in the main text. The data including this background are shown for 7.7 to 8.2\,T in Figs.~\ref{fig4}a,b. The data in the plot are shifted for clarity.} 
	\label{fig4}      
\end{figure}

Next, we turn our attention to the high-field section of the phase diagram, in which a putative phase B has been suggested on the basis of thermal expansion~\cite{stingl2013electronic}, transport~\cite{bruin2013study} and neutron scattering measurements~\cite{lester2015field}.  Previous work had not revealed any pronounced specific heat or magnetization signatures of entry to the B phase motivating in part our more detailed investigation presented here. The likely $H-T$ dependence of the phase boundary meant that the best choice near the junction with phase A would be specific heat measurements as a function of temperature at a series of closely spaced fields, while the high field part would probably be better studied with magnetocaloric traces at fixed temperatures. In Fig.~\ref{fig4}a we show the electronic specific coefficient $\gamma$ as a function of temperature from 4 to below 0.1\,K at 7.7, 7.75, 7.9 and 7.95\,T. The data for the latter two fields, cooling into the A phase, are qualitatively similar to those previously published in Refs.~\cite{rost2009entropy,rost2011thermodynamics} for temperatures above 0.25\,K. For comparison, Fig.~\ref{fig4}b shows data at the higher fields of 8.05, 8.1, 8.2, 8.3 and 8.4\,T. In the high temperature range a logarithmic divergence associated with the critical fluctuations above the A phase is observed. Then a relatively sharp kink is seen before, at low temperatures, the data fall off weakly (see also Fig.~\ref{fig3_appendix} of the Appendix). The location $T_{B}$ of this kink in $\Delta C/T$ is close to that of the expected boundary for the proposed B phase, so it is natural to speculate that $T_{B}$ is a phase transition temperature. However, there is an important qualitative difference from the feature $T_{A}$ seen at the boundary of the A phase in that the feature rapidly broadens with increasing field. Even by 8.2\,T, where the characteristic temperature of the turnover in $\Delta C/T$ remains very close to that at 8.05\,T, the broadened maximum would be difficult to associate with a phase boundary if it were seen in isolation. In the higher field section, the magnetocaloric data is even more tentative in nature.  A very weak inflection point $H_{3}$ is observed, which is coincident within experimental resolution with the feature seen in the specific heat sweeps at 8.2 and 8.3\,T. This inflection point is temperature dependent, and can be followed in magnetocaloric sweeps at all bath temperatures down to 0.05\,K, but it is a weak and broad feature at all these temperatures.

In Ref.~\cite{rost2009entropy} an empirical analysis method was proposed for the temperature-dependent $C/T$ data in the vicinity of the A phase.  The high temperature logarithmic divergence at 7.9\,T was taken to be the background of the `normal' state above $T_{c}$, extrapolated to low temperature and subtracted, in order to estimate the contribution of the phase formation to the full signal. In Fig.~\ref{fig4}d, we show the result of applying this analysis procedure to the data from this study (see also Fig.~\ref{fig3_appendix}c of the Appendix). For fields below those defining the A phase (7.7\,T), a pronounced but broad peak is seen even though there is no evidence for the development of an order parameter. The data at 7.9\,T are slightly sharper with a sharp increase below 1.1\,K suggesting a phase transition into phase A, but, interestingly, are not qualitatively different to those at 7.7 or 7.75\,T. At 8.1\,T, near the first-order transition line at $H_{2}$, a weak but sharp feature is still observed. In contrast, at 8.2\,T the main observed feature is simply a sharp drop in $\Delta C/T$ without any pronounced peak, consistent with the profound difference between the phase transitions upon entering phase A or B discussed above. One of the reasons that this subtraction method was introduced in Ref.~\cite{rost2009entropy} was to check for entropy balance associated with the A phase.  In the current work, that procedure shows entropy balance being achieved at the onset of the rise in $\Delta C/T$ within experimental error at 7.7, 7.75 (fields below those at which the A phase is thought to form) and 7.9\,T (within the A phase). It is not achieved at 8.1 or 8.2\,T, on entry to the proposed B phase. This analysis of specific heat data alone certainly cannot be used to rule out B phase formation, because it is based on an assumption about background subtraction that may not be correct.  However, Fig.~\ref{fig4}d indicates that there is not a universal functional form to the specific heat as one traverses the region of proposed phase formation.
\section{Discussion}
The QCP at 7.5\,T is of interest not just because of its position in the phase diagram, but also because it appears to be the result of a `second stage' approach to quantum criticality.  In disordered samples for $H \parallel c$, a single QCP is seen at 7.9\,T, resulting from the suppression to low temperatures of spectral weight associated, in zero field, with a 10 K energy scale~\cite{rost2011thermodynamics}.  This physics is illustrated by the specific heat and entropy data of Fig.~\ref{fig2}, and by the white symbols $T^{*}_{1}$ of Fig.~\ref{fig1} which show that the primary energy scale has been depressed to 2.5\,K by the time the applied field has reached 7\,T. Previous work has established that in more ordered samples with residual resistivity below 1\,$\mu\Omega$cm, phase A can be observed. In crystals of that quality, the single metamagnetic transition also splits into three, at 7.5, 7.8 and 8.1\,T (labeled in Fig.~\ref{fig1} $H_{0}$, $H_{1}$ and $H_{2}$ respectively).  Extensive study has shown that the higher field transitions coincide with the first-order boundaries of the A phase as identified in Fig.~\ref{fig1}, but less attention had been paid to the metamagnetic feature at $\mu_{0}H_{0} = 7.5$\,T. The data of Fig.~\ref{fig2}a show that, in these high purity samples, the 7.5\,T metamagnetic feature is consistent with the suppression to zero temperature of states associated with a second energy scale. This new scale, of as yet unknown origin, is much lower than the original one, but the qualitative thermodynamic phenomena look the same in both cases: a peak in the specific heat is depressed to progressively lower temperatures, becoming sharper and eventually producing the logarithmic divergence~\cite{lohneysen2007fermi}. 

The nature of the new 7.5\,T QCP has been suggested to be that of a 2D metamagnetic QCEP~\cite{tokiwa2016multiple} based on scaling of the magnetic Gr\"{u}neisen parameter and a fit of the electronic specific heat coefficient of the form $\Delta\gamma(H) = \gamma_{0} + \alpha(\mu_{0}(H_{c}-H))^{n}$ with fixed exponent $n = -1/3$ given by the theory \cite{millis2002metamagnetic,zacharias2013quantum}.  This fit (reproducible on our data, blue line in Fig.~\ref{fig2}b) yields $\mu_{0}H_{c}= 7.62$\,T but a value for $\gamma_0 = 58$\,mJ/Ru-molK$^2$ which is approximately half of the Sommerfeld coefficient measured at $B = 0$. In contrast, an assumption free fit to our data leaving all parameters $\gamma_0$, $\alpha$ and $n$ free and ranging from 0 to 7.4\,T (i.e. within the range in which Fermi liquid behavior is observed at low $T$) yields $n \approx -1$, $\mu_{0}H_{c}\approx 8$\,T and a more correct value for $\gamma_0 = 93$\,mJ/Ru-molK$^2$ (magenta line in Fig.~\ref{fig2}b), as was already observed in the data of Ref.~\cite{rost2011thermodynamics}. Technically, this discrepancy arises because a large change to $n$ can be compensated by a large change in $\gamma_{0}$ if the fit is carried out over a limited range of magnetic field.  More broadly, it seems dangerous to infer too much about critical scaling from fits to the field dependence of the low temperature specific heat in a system such as \sro\ in which there is the possibility of more than one critical point and in which the background density of states is likely to be strongly field dependent due to field-induced changes of the Fermi surface. It is difficult to separate these non-critical contributions to the specific heat or magnetic Gr\"{u}neisen ratio from the true critical ones. 

These concerns are somewhat less relevant for the temperature approach to criticality because determining the functional form from measurements at fixed field does not suffer from the problem of subtracting the contribution from a field-dependent background electronic structure. The detailed measurements of the temperature dependence of specific heat such as we report here are therefore of considerable relevance to understanding the nature of the criticality. In the entire region from 7.5 to 8.5\,T, specific heat data taken at fixed field over more than an order of magnitude in temperature show logarithmic rather than power law divergences over more than an order of magnitude in $T$.  Critical scaling implies that logarithmic divergences as a function of temperature go hand in hand with logarithmic divergences as a function of field, so there is a fundamental incompatibility between the temperature and field-dependent data. However, considering multiple QCPs near 8\,T, temperature scans at the critical field of a selected QCP will pick up mostly fluctuations of this particular QCP, and less those of other QCPs nearby, giving a more precise information about the nature of this particular QCP. Since the observed logarithmic divergences are weaker than power law divergences would be, our data seem compatible only with critical theories for which the dimension $d$ is equal to the dynamic critical exponent $z$, in contrast to the conclusion drawn in Ref.~\cite{tokiwa2016multiple}.

The observations that we report on `phase B' are equally intriguing. The induced transport anisotropy experiments of \cite{bruin2013study} gave a first indication that phase B is in some senses less distinctly observable than phase A, since although the rise in anisotropy at `$T_c$' was nearly as sharp as that in phase A, the level of induced anisotropy did not saturate down to the lowest temperature of measurement. Overall, the thermodynamic data reported here give only weak indications of the existence of an equilibrium phase, and do not look like the expectation for a simple transition to a spin density wave phase that gaps out part of the Fermi surface. In particular the thermodynamic signature upon entering phase A and B from high temperatures are distinctly different. This is in strong contrast to other systems where magnetic order changes via first order transitions as a function of magnetic field such as CeAuSb$_{2}$~ \cite{zhao2016field}. While this material also shows several magnetic low temperature phases separated by first order transitions, the thermodynamic features as a function of temperature such as the jump in $C/T$ are comparable in strength. Taken at face value the difference in thermodynamic properties of the A and B phase seem to indicate a fundamental difference in the phase formation and role in fluctuations of the two phases despite the seemingly similar magnetic ordering observed in neutron scattering.  Weak thermodynamic signatures reminiscent of those observed for phase B have for example been observed in another quantum critical itinerant system, NbFe$_{2}$, at the transition into a $q \neq 0$ order below about 10\,K~\cite{moroni2009magnetism}. However, it is difficult to compare the signatures observed in both systems since the NbFe$_{2}$ samples were of much worse quality than the \sro\ samples investigated here and at 10 K the specific heat of NbFe$_{2}$ is dominated by the phononic contribution. It is possible that even in \sro\ samples of the quality studied here, the disorder levels are still sufficiently high to be weakening the thermodynamic signatures of B phase formation. Moreover, the fact that there seems not to be a universal functional form to the specific heat as one traverses the region of proposed phase formation (highlighted in Fig.~\ref{fig4}d) is surprising, given the qualitative similarity of the neutron data for the proposed static order in the A and B phases.  Even more surprisingly, there is less qualitative difference between the specific heat data for the A phase and fields just below it (where no incommensurate neutron signal is seen) than there is between the proposed A and B phases.    
 
The above considerations raise important questions about the low temperature phase formation in \sro. Might the true phase diagram consist of an A phase between 7.5\,T and $H_{1}$, with $q = 0$ order but no incommensurate order, followed by A and B phases but with thermodynamic signatures that are difficult to interpret because the current levels of disorder lead to glassiness rather than true long range order? These are excellent samples, as evidenced by the prominent magnetocaloric quantum oscillations seen in Fig.~\ref{fig2}e, but they may still not be clean enough to allow the full development of fragile, disorder-dependent states.  Another concern with respect to the incommensurate magnetic order parameter might be time-scale.  Although the reported experiment in Ref.~\cite{lester2015field} is static on the 4\,$\mu$eV, 1\,GHz energy and frequency scale of the neutron scattering measurement, nuclear magnetic resonance measurements sensitive to orders of magnitude lower characteristic frequencies did not detect a relaxation time divergence at the A phase boundary~\cite{kitagawa2005metamagnetic}. These apparent differences between probes on different time scales and thermodynamic data indicate the importance of fluctuations in determining the low temperature phase diagram of \sro, an issue that clearly merits further experimental and theoretical attention. It will also be interesting to examine the extent to which realistic models for \sro, such as those discussed in Refs.~\cite{yamase2007van,puetter2010microscopic,lee2009theory,raghu2009microscopic}, contain features relating to the new data that we have presented here.
\section{Acknowledgements}
We thank Eduardo Fradkin, Markus Garst, Philipp Gegenwart, Stephen Julian, Stephen Hayden and Steven Kivelson for insightful comments, Thomas L\"uhmann for technical assistance and Philipp Gegenwart for sharing the results of magnetocaloric effect experiments performed by his group. Funding: This work was supported by the Engineering and Physical Sciences Research Council, UK (grant EP/F044704/1) and the Max Planck Society.
\appendix*
\section{APPENDIX}
This appendix contains some additional data and analysis that might be helpful for a better understanding of the paper. Since our work had the principal target to study magnetocaloric and specific heat below 0.2\,K, quantum oscillations were not carefully analysed. They are a confirmation of the good quality of the sample, but the resolution of the data for such analysis is not comparable with that in specific set-ups like those used in Refs.~\cite{mercure2010quantum,mercure2008}.

\begin{figure}[ht!]
	\begin{center}
		\includegraphics[width=\columnwidth]{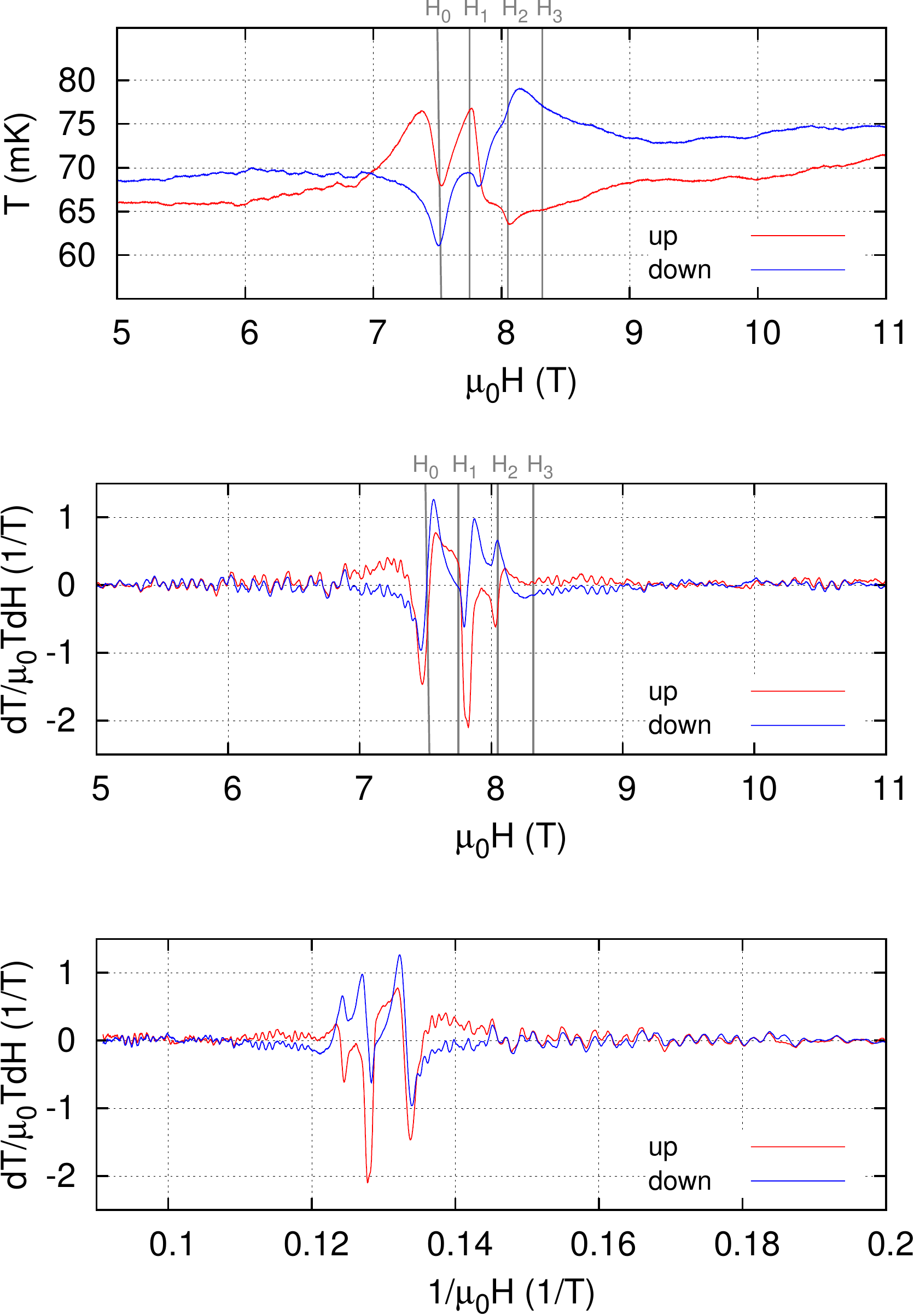}
	\end{center}
	\caption{In the upper panel we show exemplary raw data of magnetocaloric sweeps between 5 and 11\,T using the quasi-adiabatic method. Strong signatures can be seen at $H_0$, $H_1$ and $H_2$, and a weak kink at $H_3$ emphasized in Fig.~\ref{fig4}c. Below 7\,T and above 9\,T, i.e. outside the critical region, quantum oscillations are visible on this data. In the middle panel we show the field derivative of the same data, which are plotted over $1/H$ in the lower panel. 
	}
	\label{fig1_appendix}       
\end{figure}

We show in Fig.~\ref{fig1_appendix} exemplary raw data of magnetocaloric sweeps between 5 and 11\,T using the quasi-adiabatic method. In this run, we started sweeping up at a temperature of 66\,mK and back from 12\,T at a temperature of 75\,mK. All data used for the analysis in the main text had the same starting temperature in up and down sweeps. In the upper panel, we have plotted the raw data. It can immediately be seen that the sample temperature changes strongly inside the critical region, i.e. $7 \leq \mu_{0}H \leq 9$\,T, compared to the smooth change outside this region. We observe strong signatures at $H_0$, $H_1$ and $H_2$, and a weak kink at $H_3$ emphasized in Fig.~\ref{fig4}c. The strong dip at $H_{0} = 7.5$\,T is reversible while signatures between $H_{1}$ and $H_{2}$ are more complex to be understood due to possible irreversible heat introduced by friction between the sample and its mounting plate and glue, due to large magnetostriction in the region. The weak signature at $H_{3}$ is interesting: While entering the B phase from high fields (blue curve, sweep down), the sample temperature increases indicating that the entropy of the B phase is lower than that of the higher field phase. But this changes dramatically below $H_{2}$, i.e. entering the A phase, below which the temperature steeply decreases, pointing to a high-entropy A phase~\cite{rost2009entropy}.

\begin{figure}[ht!]
	\begin{center}
		\includegraphics[width=\columnwidth]{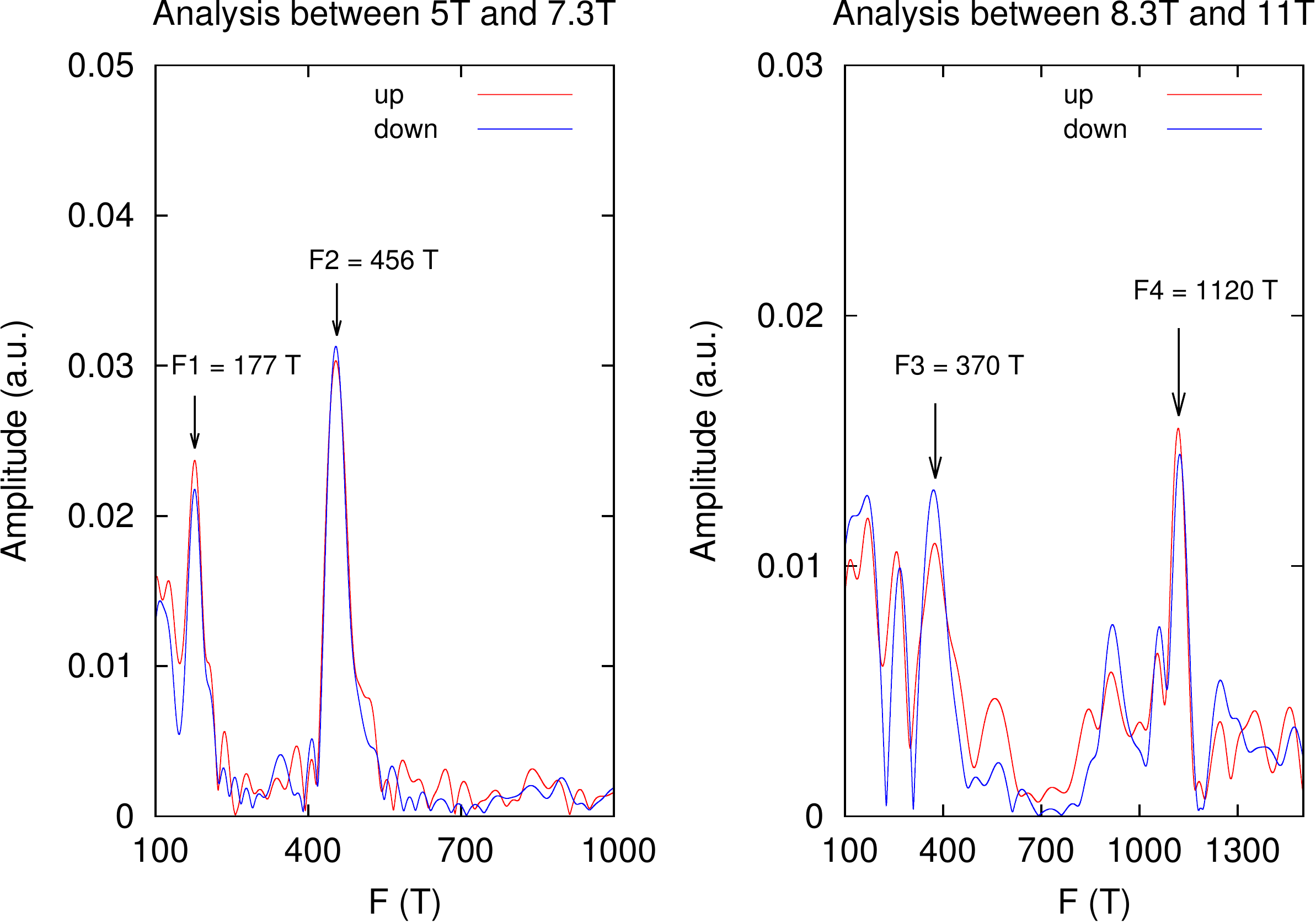}
	\end{center}
	\caption{Fourier analysis of the data shown in Fig.~\ref{fig1_appendix}. The main frequencies are the 456\,T in the low-field region and 1120\,T in the high-field region.
	}
	\label{fig2_appendix}       
\end{figure}

Below 7\,T and above 9\,T, i.e. outside the critical region, quantum oscillations are visible on this data. In the middle panel we show the field derivative of the data of the upper panel, which are plotted over $1/H$ in the lower panel. The repeatable oscillations are the result of entropy oscillations with a main frequency of about 450\,T below 7\,T (see Fig.~\ref{fig2_appendix}), already observed in magnetocaloric effect measurements~\cite{mercure2010quantum}, and a main frequency of 1120\,T above 9\,T. These oscillations are either absent or of much smaller amplitude between $H_{0}$ and $H_{3}$. Taken at face value, this might indicate a Lifshitz transition involving part of the Fermi surface, but this hypothesis needs to be checked further. As mentioned above, our set-up does not have the sensitivity to allow a proper analysis of the oscillations.
\begin{figure}[b]
	\begin{center}
		\includegraphics[width=\columnwidth]{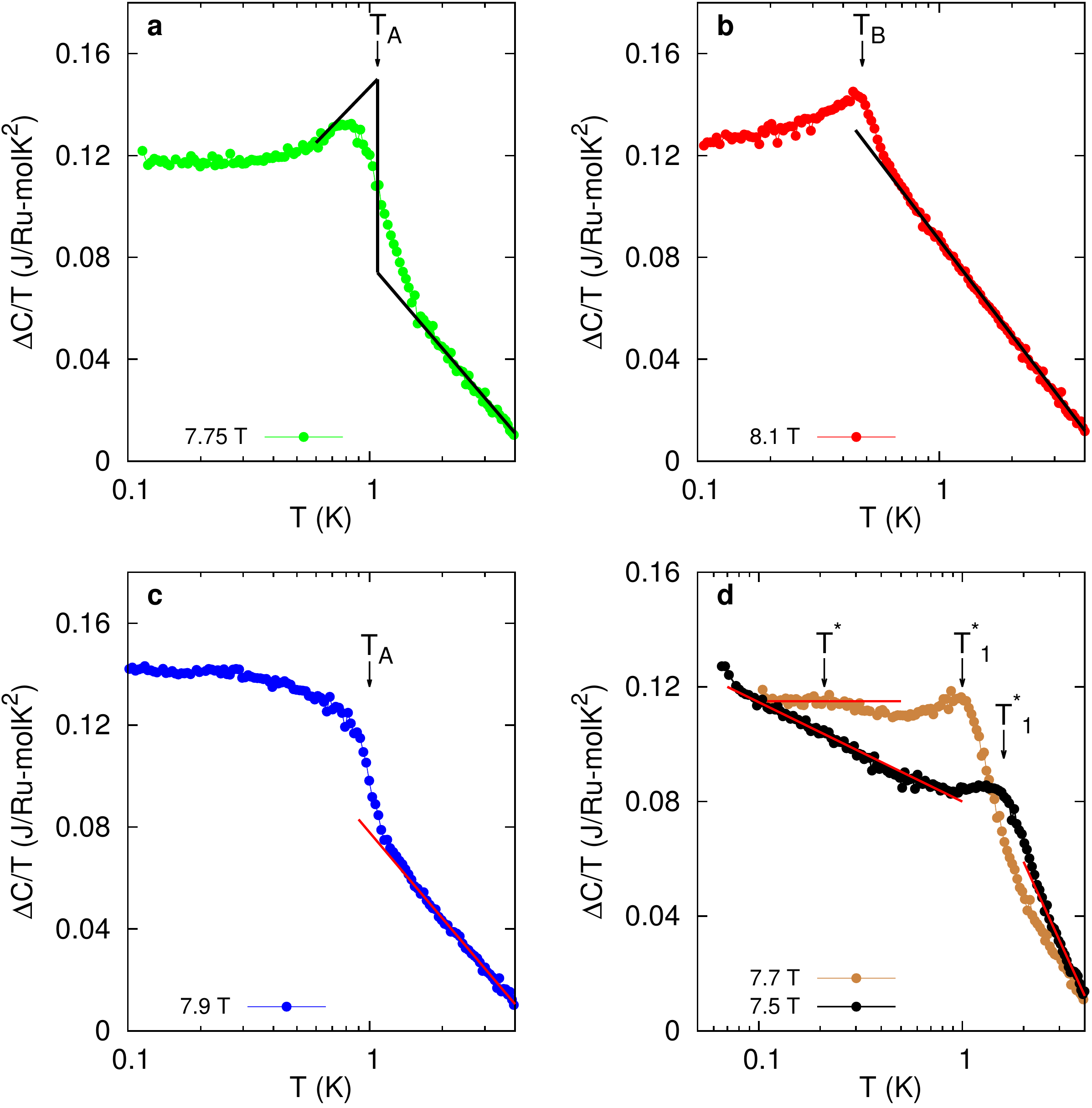}
	\end{center}
	\caption{Selected measurements to illustrate in more detail how the points in the phase diagram of Fig.~\ref{fig1} have been determined. The lines in panels (b) and (c) mark the logarithmic background that has been subtracted to obtain Fig.~\ref{fig4}d. Panel (d) emphasizes the strong logarithmic divergence observed at $H_{0} = 7.5$\,T for $T < T^{*}_{1}$.
	}
	\label{fig3_appendix}       
\end{figure}
To illustrate in more detail how we have determined the points of the phase diagram of Fig.~\ref{fig1}, we plot in Fig.~\ref{fig3_appendix} some selected measurements. We have chosen for instance the data at 7.75\,T, plotted in panel (a), which represent the transition into the A phase. While entering the A phase from high temperatures, $\Delta C/T$ deviates from the high-$T$ logarithmic behavior showing a sharp increase, a maximum and then a flattening to constant values at low-$T$. This reminds us of the signature of second order phase transitions and we used a sort of mean-field equal-entropy construction to estimate $T_{A}$. The error bar is the width of this construction. This is, however, not possible at 7.9\,T, see panel (c). Below $T_{A}$, $\Delta C/T$ still increases and in this case we estimated the center of the 'jump' in $\Delta C/T$ at $T_{A}$ and its width. While entering the B phase the situation is somehow different. The high-$T$ logarithmic behavior persists almost down to $T_{B}$ at which a sharp kink in $\Delta C/T$ is seen. Below $T_{B}$, $\Delta C/T$ decreases slowly without saturating. This is emphasized in Fig.~\ref{fig4}b and discussed in the main text. Finally, panel (d) shows the measurement at the QCP at $H_{0} = 7.5$\,T to emphasize the strong logarithmic divergence observed for $T < T^{*}_{1}$. We have also plotted the data at 7.7\,K to show how we have determined $T^{*}$ in this data, which is the point at which $\Delta C/T$ becomes constant.
\bibliography{sun_PRB}

\begin{thebibliography}{43}
\expandafter\ifx\csname natexlab\endcsname\relax\def\natexlab#1{#1}\fi
\expandafter\ifx\csname bibnamefont\endcsname\relax
  \def\bibnamefont#1{#1}\fi
\expandafter\ifx\csname bibfnamefont\endcsname\relax
  \def\bibfnamefont#1{#1}\fi
\expandafter\ifx\csname citenamefont\endcsname\relax
  \def\citenamefont#1{#1}\fi
\expandafter\ifx\csname url\endcsname\relax
  \def\url#1{\texttt{#1}}\fi
\expandafter\ifx\csname urlprefix\endcsname\relax\def\urlprefix{URL }\fi
\providecommand{\bibinfo}[2]{#2}
\providecommand{\eprint}[2][]{\url{#2}}

\bibitem[{\citenamefont{Cao et~al.}(1997)\citenamefont{Cao, McCall, and
  Crow}}]{cao1997observation}
\bibinfo{author}{\bibfnamefont{G.}~\bibnamefont{Cao}},
  \bibinfo{author}{\bibfnamefont{S.}~\bibnamefont{McCall}}, \bibnamefont{and}
  \bibinfo{author}{\bibfnamefont{J.~E.} \bibnamefont{Crow}},
  \bibinfo{journal}{Phys. Rev. B} \textbf{\bibinfo{volume}{55}},
  \bibinfo{pages}{R672} (\bibinfo{year}{1997}),
  \urlprefix\url{https://link.aps.org/doi/10.1103/PhysRevB.55.R672}.

\bibitem[{\citenamefont{Ikeda et~al.}(2000)\citenamefont{Ikeda, Maeno,
  Nakatsuji, Kosaka, and Uwatoko}}]{ikeda2000ground}
\bibinfo{author}{\bibfnamefont{S.-I.} \bibnamefont{Ikeda}},
  \bibinfo{author}{\bibfnamefont{Y.}~\bibnamefont{Maeno}},
  \bibinfo{author}{\bibfnamefont{S.}~\bibnamefont{Nakatsuji}},
  \bibinfo{author}{\bibfnamefont{M.}~\bibnamefont{Kosaka}}, \bibnamefont{and}
  \bibinfo{author}{\bibfnamefont{Y.}~\bibnamefont{Uwatoko}},
  \bibinfo{journal}{Phys. Rev. B} \textbf{\bibinfo{volume}{62}},
  \bibinfo{pages}{R6089} (\bibinfo{year}{2000}),
  \urlprefix\url{https://link.aps.org/doi/10.1103/PhysRevB.62.R6089}.

\bibitem[{\citenamefont{Mackenzie et~al.}(2012)\citenamefont{Mackenzie, Bruin,
  Borzi, Rost, and Grigera}}]{mackenzie2012quantum}
\bibinfo{author}{\bibfnamefont{A.~P.} \bibnamefont{Mackenzie}},
  \bibinfo{author}{\bibfnamefont{J.~A.~N.} \bibnamefont{Bruin}},
  \bibinfo{author}{\bibfnamefont{R.~A.} \bibnamefont{Borzi}},
  \bibinfo{author}{\bibfnamefont{A.~W.} \bibnamefont{Rost}}, \bibnamefont{and}
  \bibinfo{author}{\bibfnamefont{S.~A.} \bibnamefont{Grigera}},
  \bibinfo{journal}{Phys. C} \textbf{\bibinfo{volume}{481}},
  \bibinfo{pages}{207} (\bibinfo{year}{2012}).

\bibitem[{\citenamefont{Perry and Maeno}(2004)}]{perry2004systematic}
\bibinfo{author}{\bibfnamefont{R.~S.} \bibnamefont{Perry}} \bibnamefont{and}
  \bibinfo{author}{\bibfnamefont{Y.}~\bibnamefont{Maeno}}, \bibinfo{journal}{J.
  Cryst. Growth} \textbf{\bibinfo{volume}{271}}, \bibinfo{pages}{134}
  (\bibinfo{year}{2004}).

\bibitem[{\citenamefont{Grigera et~al.}(2003)\citenamefont{Grigera, Borzi,
  Mackenzie, Julian, Perry, and Maeno}}]{grigera2003angular}
\bibinfo{author}{\bibfnamefont{S.~A.} \bibnamefont{Grigera}},
  \bibinfo{author}{\bibfnamefont{R.~A.} \bibnamefont{Borzi}},
  \bibinfo{author}{\bibfnamefont{A.~P.} \bibnamefont{Mackenzie}},
  \bibinfo{author}{\bibfnamefont{S.~R.} \bibnamefont{Julian}},
  \bibinfo{author}{\bibfnamefont{R.~S.} \bibnamefont{Perry}}, \bibnamefont{and}
  \bibinfo{author}{\bibfnamefont{Y.}~\bibnamefont{Maeno}},
  \bibinfo{journal}{Phys. Rev. B} \textbf{\bibinfo{volume}{67}},
  \bibinfo{pages}{214427} (\bibinfo{year}{2003}),
  \urlprefix\url{https://link.aps.org/doi/10.1103/PhysRevB.67.214427}.

\bibitem[{\citenamefont{Grigera et~al.}(2004)\citenamefont{Grigera, Gegenwart,
  Borzi, Weickert, Schofield, Perry, Tayama, Sakakibara, Maeno, Green
  et~al.}}]{grigera2004disorder}
\bibinfo{author}{\bibfnamefont{S.~A.} \bibnamefont{Grigera}},
  \bibinfo{author}{\bibfnamefont{P.}~\bibnamefont{Gegenwart}},
  \bibinfo{author}{\bibfnamefont{R.~A.} \bibnamefont{Borzi}},
  \bibinfo{author}{\bibfnamefont{F.}~\bibnamefont{Weickert}},
  \bibinfo{author}{\bibfnamefont{A.~J.} \bibnamefont{Schofield}},
  \bibinfo{author}{\bibfnamefont{R.~S.} \bibnamefont{Perry}},
  \bibinfo{author}{\bibfnamefont{T.}~\bibnamefont{Tayama}},
  \bibinfo{author}{\bibfnamefont{T.}~\bibnamefont{Sakakibara}},
  \bibinfo{author}{\bibfnamefont{Y.}~\bibnamefont{Maeno}},
  \bibinfo{author}{\bibfnamefont{A.~G.} \bibnamefont{Green}},
  \bibnamefont{et~al.}, \bibinfo{journal}{Science}
  \textbf{\bibinfo{volume}{306}}, \bibinfo{pages}{1154} (\bibinfo{year}{2004}).

\bibitem[{\citenamefont{Rost et~al.}(2009)\citenamefont{Rost, Perry, Mercure,
  Mackenzie, and Grigera}}]{rost2009entropy}
\bibinfo{author}{\bibfnamefont{A.~W.} \bibnamefont{Rost}},
  \bibinfo{author}{\bibfnamefont{R.~S.} \bibnamefont{Perry}},
  \bibinfo{author}{\bibfnamefont{J.-F.} \bibnamefont{Mercure}},
  \bibinfo{author}{\bibfnamefont{A.~P.} \bibnamefont{Mackenzie}},
  \bibnamefont{and} \bibinfo{author}{\bibfnamefont{S.~A.}
  \bibnamefont{Grigera}}, \bibinfo{journal}{Science}
  \textbf{\bibinfo{volume}{325}}, \bibinfo{pages}{1360} (\bibinfo{year}{2009}).

\bibitem[{\citenamefont{Gegenwart et~al.}(2006)\citenamefont{Gegenwart,
  Weickert, Garst, Perry, and Maeno}}]{gegenwart2006metamagnetic}
\bibinfo{author}{\bibfnamefont{P.}~\bibnamefont{Gegenwart}},
  \bibinfo{author}{\bibfnamefont{F.}~\bibnamefont{Weickert}},
  \bibinfo{author}{\bibfnamefont{M.}~\bibnamefont{Garst}},
  \bibinfo{author}{\bibfnamefont{R.~S.} \bibnamefont{Perry}}, \bibnamefont{and}
  \bibinfo{author}{\bibfnamefont{Y.}~\bibnamefont{Maeno}},
  \bibinfo{journal}{Phys. Rev. Lett.} \textbf{\bibinfo{volume}{96}},
  \bibinfo{pages}{136402} (\bibinfo{year}{2006}),
  \urlprefix\url{https://link.aps.org/doi/10.1103/PhysRevLett.96.136402}.

\bibitem[{\citenamefont{Stingl et~al.}(2013)\citenamefont{Stingl, Perry, Maeno,
  and Gegenwart}}]{stingl2013electronic}
\bibinfo{author}{\bibfnamefont{C.}~\bibnamefont{Stingl}},
  \bibinfo{author}{\bibfnamefont{R.~S.} \bibnamefont{Perry}},
  \bibinfo{author}{\bibfnamefont{Y.}~\bibnamefont{Maeno}}, \bibnamefont{and}
  \bibinfo{author}{\bibfnamefont{P.}~\bibnamefont{Gegenwart}},
  \bibinfo{journal}{Phys. Status Solidi (b)} \textbf{\bibinfo{volume}{250}},
  \bibinfo{pages}{450} (\bibinfo{year}{2013}).

\bibitem[{\citenamefont{Bruin et~al.}(2013)\citenamefont{Bruin, Borzi, Grigera,
  Rost, Perry, and Mackenzie}}]{bruin2013study}
\bibinfo{author}{\bibfnamefont{J.~A.~N.} \bibnamefont{Bruin}},
  \bibinfo{author}{\bibfnamefont{R.~A.} \bibnamefont{Borzi}},
  \bibinfo{author}{\bibfnamefont{S.~A.} \bibnamefont{Grigera}},
  \bibinfo{author}{\bibfnamefont{A.~W.} \bibnamefont{Rost}},
  \bibinfo{author}{\bibfnamefont{R.~S.} \bibnamefont{Perry}}, \bibnamefont{and}
  \bibinfo{author}{\bibfnamefont{A.~P.} \bibnamefont{Mackenzie}},
  \bibinfo{journal}{Phys. Rev. B} \textbf{\bibinfo{volume}{87}},
  \bibinfo{pages}{161106} (\bibinfo{year}{2013}),
  \urlprefix\url{https://link.aps.org/doi/10.1103/PhysRevB.87.161106}.

\bibitem[{\citenamefont{Lester et~al.}(2015)\citenamefont{Lester, Ramos, Perry,
  Croft, Bewley, Guidi, Manuel, Khalyavin, Forgan, and
  Hayden}}]{lester2015field}
\bibinfo{author}{\bibfnamefont{C.}~\bibnamefont{Lester}},
  \bibinfo{author}{\bibfnamefont{S.}~\bibnamefont{Ramos}},
  \bibinfo{author}{\bibfnamefont{R.~S.} \bibnamefont{Perry}},
  \bibinfo{author}{\bibfnamefont{T.~P.} \bibnamefont{Croft}},
  \bibinfo{author}{\bibfnamefont{R.~I.} \bibnamefont{Bewley}},
  \bibinfo{author}{\bibfnamefont{T.}~\bibnamefont{Guidi}},
  \bibinfo{author}{\bibfnamefont{P.}~\bibnamefont{Manuel}},
  \bibinfo{author}{\bibfnamefont{D.~D.} \bibnamefont{Khalyavin}},
  \bibinfo{author}{\bibfnamefont{E.~M.} \bibnamefont{Forgan}},
  \bibnamefont{and} \bibinfo{author}{\bibfnamefont{S.~M.}
  \bibnamefont{Hayden}}, \bibinfo{journal}{Nat. Mater.}
  \textbf{\bibinfo{volume}{14}}, \bibinfo{pages}{373} (\bibinfo{year}{2015}).

\bibitem[{\citenamefont{Borzi et~al.}(2007)\citenamefont{Borzi, Grigera,
  Farrell, Perry, Lister, Lee, Tennant, Maeno, and
  Mackenzie}}]{borzi2007formation}
\bibinfo{author}{\bibfnamefont{R.~A.} \bibnamefont{Borzi}},
  \bibinfo{author}{\bibfnamefont{S.~A.} \bibnamefont{Grigera}},
  \bibinfo{author}{\bibfnamefont{J.}~\bibnamefont{Farrell}},
  \bibinfo{author}{\bibfnamefont{R.~S.} \bibnamefont{Perry}},
  \bibinfo{author}{\bibfnamefont{S.~J.~S.} \bibnamefont{Lister}},
  \bibinfo{author}{\bibfnamefont{S.~L.} \bibnamefont{Lee}},
  \bibinfo{author}{\bibfnamefont{D.~A.} \bibnamefont{Tennant}},
  \bibinfo{author}{\bibfnamefont{Y.}~\bibnamefont{Maeno}}, \bibnamefont{and}
  \bibinfo{author}{\bibfnamefont{A.~P.} \bibnamefont{Mackenzie}},
  \bibinfo{journal}{Science} \textbf{\bibinfo{volume}{315}},
  \bibinfo{pages}{214} (\bibinfo{year}{2007}).

\bibitem[{\citenamefont{Brodsky et~al.}(2017)\citenamefont{Brodsky, Barber,
  Bruin, Borzi, Grigera, Perry, Mackenzie, and Hicks}}]{brodsky2017strain}
\bibinfo{author}{\bibfnamefont{D.~O.} \bibnamefont{Brodsky}},
  \bibinfo{author}{\bibfnamefont{M.~E.} \bibnamefont{Barber}},
  \bibinfo{author}{\bibfnamefont{J.~A.~N.} \bibnamefont{Bruin}},
  \bibinfo{author}{\bibfnamefont{R.~A.} \bibnamefont{Borzi}},
  \bibinfo{author}{\bibfnamefont{S.~A.} \bibnamefont{Grigera}},
  \bibinfo{author}{\bibfnamefont{R.~S.} \bibnamefont{Perry}},
  \bibinfo{author}{\bibfnamefont{A.~P.} \bibnamefont{Mackenzie}},
  \bibnamefont{and} \bibinfo{author}{\bibfnamefont{C.~W.} \bibnamefont{Hicks}},
  \bibinfo{journal}{Sci. Adv.} \textbf{\bibinfo{volume}{3}},
  \bibinfo{pages}{e1501804} (\bibinfo{year}{2017}).

\bibitem[{\citenamefont{Hase and Nishihara}(1997)}]{hase1997electronic}
\bibinfo{author}{\bibfnamefont{I.}~\bibnamefont{Hase}} \bibnamefont{and}
  \bibinfo{author}{\bibfnamefont{Y.}~\bibnamefont{Nishihara}},
  \bibinfo{journal}{J. Phys. Soc. Jpn.} \textbf{\bibinfo{volume}{66}},
  \bibinfo{pages}{3517} (\bibinfo{year}{1997}).

\bibitem[{\citenamefont{Singh and Mazin}(2001)}]{singh2001electronic}
\bibinfo{author}{\bibfnamefont{D.~J.} \bibnamefont{Singh}} \bibnamefont{and}
  \bibinfo{author}{\bibfnamefont{I.~I.} \bibnamefont{Mazin}},
  \bibinfo{journal}{Phys. Rev. B} \textbf{\bibinfo{volume}{63}},
  \bibinfo{pages}{165101} (\bibinfo{year}{2001}),
  \urlprefix\url{https://link.aps.org/doi/10.1103/PhysRevB.63.165101}.

\bibitem[{\citenamefont{Tamai et~al.}(2008)\citenamefont{Tamai, Allan, Mercure,
  Meevasana, Dunkel, Lu, Perry, Mackenzie, Singh, Shen
  et~al.}}]{tamai2008fermi}
\bibinfo{author}{\bibfnamefont{A.}~\bibnamefont{Tamai}},
  \bibinfo{author}{\bibfnamefont{M.~P.} \bibnamefont{Allan}},
  \bibinfo{author}{\bibfnamefont{J.~F.} \bibnamefont{Mercure}},
  \bibinfo{author}{\bibfnamefont{W.}~\bibnamefont{Meevasana}},
  \bibinfo{author}{\bibfnamefont{R.}~\bibnamefont{Dunkel}},
  \bibinfo{author}{\bibfnamefont{D.~H.} \bibnamefont{Lu}},
  \bibinfo{author}{\bibfnamefont{R.~S.} \bibnamefont{Perry}},
  \bibinfo{author}{\bibfnamefont{A.~P.} \bibnamefont{Mackenzie}},
  \bibinfo{author}{\bibfnamefont{D.~J.} \bibnamefont{Singh}},
  \bibinfo{author}{\bibfnamefont{Z.-X.} \bibnamefont{Shen}},
  \bibnamefont{et~al.}, \bibinfo{journal}{Phys. Rev. Lett.}
  \textbf{\bibinfo{volume}{101}}, \bibinfo{pages}{026407}
  (\bibinfo{year}{2008}),
  \urlprefix\url{https://link.aps.org/doi/10.1103/PhysRevLett.101.026407}.

\bibitem[{\citenamefont{Allan et~al.}(2013)\citenamefont{Allan, Tamai,
  Rozbicki, Fischer, Voss, King, Meevasana, Thirupathaiah, Rienks, Fink
  et~al.}}]{allan2013formation}
\bibinfo{author}{\bibfnamefont{M.~P.} \bibnamefont{Allan}},
  \bibinfo{author}{\bibfnamefont{A.}~\bibnamefont{Tamai}},
  \bibinfo{author}{\bibfnamefont{E.}~\bibnamefont{Rozbicki}},
  \bibinfo{author}{\bibfnamefont{M.~H.} \bibnamefont{Fischer}},
  \bibinfo{author}{\bibfnamefont{J.}~\bibnamefont{Voss}},
  \bibinfo{author}{\bibfnamefont{P.~D.~C.} \bibnamefont{King}},
  \bibinfo{author}{\bibfnamefont{W.}~\bibnamefont{Meevasana}},
  \bibinfo{author}{\bibfnamefont{S.}~\bibnamefont{Thirupathaiah}},
  \bibinfo{author}{\bibfnamefont{E.}~\bibnamefont{Rienks}},
  \bibinfo{author}{\bibfnamefont{J.}~\bibnamefont{Fink}}, \bibnamefont{et~al.},
  \bibinfo{journal}{New Journal of Physics} \textbf{\bibinfo{volume}{15}},
  \bibinfo{pages}{063029} (\bibinfo{year}{2013}),
  \urlprefix\url{http://stacks.iop.org/1367-2630/15/i=6/a=063029}.

\bibitem[{\citenamefont{Borzi et~al.}(2004)\citenamefont{Borzi, Grigera, Perry,
  Kikugawa, Kitagawa, Maeno, and Mackenzie}}]{borzi2004haas}
\bibinfo{author}{\bibfnamefont{R.~A.} \bibnamefont{Borzi}},
  \bibinfo{author}{\bibfnamefont{S.~A.} \bibnamefont{Grigera}},
  \bibinfo{author}{\bibfnamefont{R.~S.} \bibnamefont{Perry}},
  \bibinfo{author}{\bibfnamefont{N.}~\bibnamefont{Kikugawa}},
  \bibinfo{author}{\bibfnamefont{K.}~\bibnamefont{Kitagawa}},
  \bibinfo{author}{\bibfnamefont{Y.}~\bibnamefont{Maeno}}, \bibnamefont{and}
  \bibinfo{author}{\bibfnamefont{A.~P.} \bibnamefont{Mackenzie}},
  \bibinfo{journal}{Phys. Rev. Lett.} \textbf{\bibinfo{volume}{92}},
  \bibinfo{pages}{216403} (\bibinfo{year}{2004}),
  \urlprefix\url{https://link.aps.org/doi/10.1103/PhysRevLett.92.216403}.

\bibitem[{\citenamefont{Mercure et~al.}(2010)\citenamefont{Mercure, Rost,
  O'Farrell, Goh, Perry, Sutherland, Grigera, Borzi, Gegenwart, Gibbs
  et~al.}}]{mercure2010quantum}
\bibinfo{author}{\bibfnamefont{J.-F.} \bibnamefont{Mercure}},
  \bibinfo{author}{\bibfnamefont{A.~W.} \bibnamefont{Rost}},
  \bibinfo{author}{\bibfnamefont{E.~C.~T.} \bibnamefont{O'Farrell}},
  \bibinfo{author}{\bibfnamefont{S.~K.} \bibnamefont{Goh}},
  \bibinfo{author}{\bibfnamefont{R.~S.} \bibnamefont{Perry}},
  \bibinfo{author}{\bibfnamefont{M.~L.} \bibnamefont{Sutherland}},
  \bibinfo{author}{\bibfnamefont{S.~A.} \bibnamefont{Grigera}},
  \bibinfo{author}{\bibfnamefont{R.~A.} \bibnamefont{Borzi}},
  \bibinfo{author}{\bibfnamefont{P.}~\bibnamefont{Gegenwart}},
  \bibinfo{author}{\bibfnamefont{A.~S.} \bibnamefont{Gibbs}},
  \bibnamefont{et~al.}, \bibinfo{journal}{Phys. Rev. B}
  \textbf{\bibinfo{volume}{81}}, \bibinfo{pages}{235103}
  (\bibinfo{year}{2010}),
  \urlprefix\url{https://link.aps.org/doi/10.1103/PhysRevB.81.235103}.

\bibitem[{\citenamefont{Tokiwa et~al.}(2016)\citenamefont{Tokiwa, Mchalwat,
  Perry, and Gegenwart}}]{tokiwa2016multiple}
\bibinfo{author}{\bibfnamefont{Y.}~\bibnamefont{Tokiwa}},
  \bibinfo{author}{\bibfnamefont{M.}~\bibnamefont{Mchalwat}},
  \bibinfo{author}{\bibfnamefont{R.~S.} \bibnamefont{Perry}}, \bibnamefont{and}
  \bibinfo{author}{\bibfnamefont{P.}~\bibnamefont{Gegenwart}},
  \bibinfo{journal}{Phys. Rev. Lett.} \textbf{\bibinfo{volume}{116}},
  \bibinfo{pages}{226402} (\bibinfo{year}{2016}),
  \urlprefix\url{https://link.aps.org/doi/10.1103/PhysRevLett.116.226402}.

\bibitem[{\citenamefont{Ikeda et~al.}(2004)\citenamefont{Ikeda, Shirakawa,
  Yanagisawa, Yoshida, Koikegami, Koike, Kosaka, and
  Uwatoko}}]{ikeda2004uniaxial}
\bibinfo{author}{\bibfnamefont{S.-I.} \bibnamefont{Ikeda}},
  \bibinfo{author}{\bibfnamefont{N.}~\bibnamefont{Shirakawa}},
  \bibinfo{author}{\bibfnamefont{T.}~\bibnamefont{Yanagisawa}},
  \bibinfo{author}{\bibfnamefont{Y.}~\bibnamefont{Yoshida}},
  \bibinfo{author}{\bibfnamefont{S.}~\bibnamefont{Koikegami}},
  \bibinfo{author}{\bibfnamefont{S.}~\bibnamefont{Koike}},
  \bibinfo{author}{\bibfnamefont{M.}~\bibnamefont{Kosaka}}, \bibnamefont{and}
  \bibinfo{author}{\bibfnamefont{Y.}~\bibnamefont{Uwatoko}},
  \bibinfo{journal}{J. Phys. Soc. Jpn.} \textbf{\bibinfo{volume}{73}},
  \bibinfo{pages}{1322} (\bibinfo{year}{2004}).

\bibitem[{\citenamefont{Perry et~al.}(2001)\citenamefont{Perry, Galvin,
  Grigera, Capogna, Schofield, Mackenzie, Chiao, Julian, Ikeda, Nakatsuji
  et~al.}}]{perry2001metamagnetism}
\bibinfo{author}{\bibfnamefont{R.~S.} \bibnamefont{Perry}},
  \bibinfo{author}{\bibfnamefont{L.~M.} \bibnamefont{Galvin}},
  \bibinfo{author}{\bibfnamefont{S.~A.} \bibnamefont{Grigera}},
  \bibinfo{author}{\bibfnamefont{L.}~\bibnamefont{Capogna}},
  \bibinfo{author}{\bibfnamefont{A.~J.} \bibnamefont{Schofield}},
  \bibinfo{author}{\bibfnamefont{A.~P.} \bibnamefont{Mackenzie}},
  \bibinfo{author}{\bibfnamefont{M.}~\bibnamefont{Chiao}},
  \bibinfo{author}{\bibfnamefont{S.~R.} \bibnamefont{Julian}},
  \bibinfo{author}{\bibfnamefont{S.~I.} \bibnamefont{Ikeda}},
  \bibinfo{author}{\bibfnamefont{S.}~\bibnamefont{Nakatsuji}},
  \bibnamefont{et~al.}, \bibinfo{journal}{Phys. Rev. Lett.}
  \textbf{\bibinfo{volume}{86}}, \bibinfo{pages}{2661} (\bibinfo{year}{2001}),
  \urlprefix\url{https://link.aps.org/doi/10.1103/PhysRevLett.86.2661}.

\bibitem[{\citenamefont{Chiao et~al.}(2002)\citenamefont{Chiao, Pfleiderer,
  Julian, Lonzarich, Perry, Mackenzie, and Maeno}}]{chiao2002effect}
\bibinfo{author}{\bibfnamefont{M.}~\bibnamefont{Chiao}},
  \bibinfo{author}{\bibfnamefont{C.}~\bibnamefont{Pfleiderer}},
  \bibinfo{author}{\bibfnamefont{S.}~\bibnamefont{Julian}},
  \bibinfo{author}{\bibfnamefont{G.}~\bibnamefont{Lonzarich}},
  \bibinfo{author}{\bibfnamefont{R.}~\bibnamefont{Perry}},
  \bibinfo{author}{\bibfnamefont{A.}~\bibnamefont{Mackenzie}},
  \bibnamefont{and} \bibinfo{author}{\bibfnamefont{Y.}~\bibnamefont{Maeno}},
  \bibinfo{journal}{Physica B: Condensed Matter}
  \textbf{\bibinfo{volume}{312--313}}, \bibinfo{pages}{698}
  (\bibinfo{year}{2002}), ISSN \bibinfo{issn}{0921--4526}, \bibinfo{note}{the
  International Conference on Strongly Correlated Electron Systems},
  \urlprefix\url{http://www.sciencedirect.com/science/article/pii/S0921452601012030}.

\bibitem[{\citenamefont{Wu et~al.}(2011)\citenamefont{Wu, McCollam, Grigera,
  Perry, Mackenzie, and Julian}}]{wu2011quantum}
\bibinfo{author}{\bibfnamefont{W.}~\bibnamefont{Wu}},
  \bibinfo{author}{\bibfnamefont{A.}~\bibnamefont{McCollam}},
  \bibinfo{author}{\bibfnamefont{S.~A.} \bibnamefont{Grigera}},
  \bibinfo{author}{\bibfnamefont{R.~S.} \bibnamefont{Perry}},
  \bibinfo{author}{\bibfnamefont{A.~P.} \bibnamefont{Mackenzie}},
  \bibnamefont{and} \bibinfo{author}{\bibfnamefont{S.~R.}
  \bibnamefont{Julian}}, \bibinfo{journal}{Phys. Rev. B}
  \textbf{\bibinfo{volume}{83}}, \bibinfo{pages}{045106}
  (\bibinfo{year}{2011}),
  \urlprefix\url{https://link.aps.org/doi/10.1103/PhysRevB.83.045106}.

\bibitem[{\citenamefont{Sun et~al.}(2013)\citenamefont{Sun, Wu, Grigera, Perry,
  Mackenzie, and Julian}}]{sun2013pressure}
\bibinfo{author}{\bibfnamefont{D.}~\bibnamefont{Sun}},
  \bibinfo{author}{\bibfnamefont{W.}~\bibnamefont{Wu}},
  \bibinfo{author}{\bibfnamefont{S.~A.} \bibnamefont{Grigera}},
  \bibinfo{author}{\bibfnamefont{R.~S.} \bibnamefont{Perry}},
  \bibinfo{author}{\bibfnamefont{A.~P.} \bibnamefont{Mackenzie}},
  \bibnamefont{and} \bibinfo{author}{\bibfnamefont{S.~R.}
  \bibnamefont{Julian}}, \bibinfo{journal}{Phys. Rev. B}
  \textbf{\bibinfo{volume}{88}}, \bibinfo{pages}{235129}
  (\bibinfo{year}{2013}),
  \urlprefix\url{https://link.aps.org/doi/10.1103/PhysRevB.88.235129}.

\bibitem[{\citenamefont{Millis et~al.}(2002)\citenamefont{Millis, Schofield,
  Lonzarich, and Grigera}}]{millis2002metamagnetic}
\bibinfo{author}{\bibfnamefont{A.~J.} \bibnamefont{Millis}},
  \bibinfo{author}{\bibfnamefont{A.~J.} \bibnamefont{Schofield}},
  \bibinfo{author}{\bibfnamefont{G.~G.} \bibnamefont{Lonzarich}},
  \bibnamefont{and} \bibinfo{author}{\bibfnamefont{S.~A.}
  \bibnamefont{Grigera}}, \bibinfo{journal}{Phys. Rev. Lett.}
  \textbf{\bibinfo{volume}{88}}, \bibinfo{pages}{217204}
  (\bibinfo{year}{2002}),
  \urlprefix\url{https://link.aps.org/doi/10.1103/PhysRevLett.88.217204}.

\bibitem[{\citenamefont{Ikeda et~al.}(2002)\citenamefont{Ikeda, Azuma,
  Shirakawa, Nishihara, and Maeno}}]{ikeda2002bulk}
\bibinfo{author}{\bibfnamefont{S.~I.} \bibnamefont{Ikeda}},
  \bibinfo{author}{\bibfnamefont{U.}~\bibnamefont{Azuma}},
  \bibinfo{author}{\bibfnamefont{N.}~\bibnamefont{Shirakawa}},
  \bibinfo{author}{\bibfnamefont{Y.}~\bibnamefont{Nishihara}},
  \bibnamefont{and} \bibinfo{author}{\bibfnamefont{Y.}~\bibnamefont{Maeno}},
  \bibinfo{journal}{J. Cryst. Growth} \textbf{\bibinfo{volume}{237}},
  \bibinfo{pages}{787} (\bibinfo{year}{2002}).

\bibitem[{\citenamefont{Wilhelm et~al.}(2004)\citenamefont{Wilhelm, L\"uhmann,
  Rus, and Steglich}}]{Wilhelm2004}
\bibinfo{author}{\bibfnamefont{H.}~\bibnamefont{Wilhelm}},
  \bibinfo{author}{\bibfnamefont{T.}~\bibnamefont{L\"uhmann}},
  \bibinfo{author}{\bibfnamefont{T.}~\bibnamefont{Rus}}, \bibnamefont{and}
  \bibinfo{author}{\bibfnamefont{F.}~\bibnamefont{Steglich}},
  \bibinfo{journal}{Rev. Sci. Instrum.} \textbf{\bibinfo{volume}{75}},
  \bibinfo{pages}{2700} (\bibinfo{year}{2004}).

\bibitem[{\citenamefont{Rost}(2009)}]{rost2009thesis}
\bibinfo{author}{\bibfnamefont{A.~W.} \bibnamefont{Rost}}, Ph.D. thesis,
  \bibinfo{school}{University of St Andrews} (\bibinfo{year}{2009}).

\bibitem[{\citenamefont{Rost et~al.}(2011)\citenamefont{Rost, Grigera, Bruin,
  Perry, Tian, Raghu, Kivelson, and Mackenzie}}]{rost2011thermodynamics}
\bibinfo{author}{\bibfnamefont{A.~W.} \bibnamefont{Rost}},
  \bibinfo{author}{\bibfnamefont{S.~A.} \bibnamefont{Grigera}},
  \bibinfo{author}{\bibfnamefont{J.~A.~N.} \bibnamefont{Bruin}},
  \bibinfo{author}{\bibfnamefont{R.~S.} \bibnamefont{Perry}},
  \bibinfo{author}{\bibfnamefont{D.}~\bibnamefont{Tian}},
  \bibinfo{author}{\bibfnamefont{S.}~\bibnamefont{Raghu}},
  \bibinfo{author}{\bibfnamefont{S.~A.} \bibnamefont{Kivelson}},
  \bibnamefont{and} \bibinfo{author}{\bibfnamefont{A.~P.}
  \bibnamefont{Mackenzie}}, \bibinfo{journal}{Proc. Natl. Acad. Sci. U.S.A.}
  \textbf{\bibinfo{volume}{108}}, \bibinfo{pages}{16549}
  (\bibinfo{year}{2011}).

\bibitem[{\citenamefont{Zhu et~al.}(2003)\citenamefont{Zhu, Garst, Rosch, and
  Si}}]{zhu2003universally}
\bibinfo{author}{\bibfnamefont{L.}~\bibnamefont{Zhu}},
  \bibinfo{author}{\bibfnamefont{M.}~\bibnamefont{Garst}},
  \bibinfo{author}{\bibfnamefont{A.}~\bibnamefont{Rosch}}, \bibnamefont{and}
  \bibinfo{author}{\bibfnamefont{Q.}~\bibnamefont{Si}}, \bibinfo{journal}{Phys.
  Rev. Lett.} \textbf{\bibinfo{volume}{91}}, \bibinfo{pages}{066404}
  (\bibinfo{year}{2003}),
  \urlprefix\url{https://link.aps.org/doi/10.1103/PhysRevLett.91.066404}.

\bibitem[{\citenamefont{Garst and Rosch}(2005)}]{garst2005sign}
\bibinfo{author}{\bibfnamefont{M.}~\bibnamefont{Garst}} \bibnamefont{and}
  \bibinfo{author}{\bibfnamefont{A.}~\bibnamefont{Rosch}},
  \bibinfo{journal}{Phys. Rev. B} \textbf{\bibinfo{volume}{72}},
  \bibinfo{pages}{205129} (\bibinfo{year}{2005}),
  \urlprefix\url{https://link.aps.org/doi/10.1103/PhysRevB.72.205129}.

\bibitem[{\citenamefont{L\"ohneysen et~al.}(2007)\citenamefont{L\"ohneysen,
  Rosch, Vojta, and W\"olfle}}]{lohneysen2007fermi}
\bibinfo{author}{\bibfnamefont{H.~v.} \bibnamefont{L\"ohneysen}},
  \bibinfo{author}{\bibfnamefont{A.}~\bibnamefont{Rosch}},
  \bibinfo{author}{\bibfnamefont{M.}~\bibnamefont{Vojta}}, \bibnamefont{and}
  \bibinfo{author}{\bibfnamefont{P.}~\bibnamefont{W\"olfle}},
  \bibinfo{journal}{Rev. Mod. Phys.} \textbf{\bibinfo{volume}{79}},
  \bibinfo{pages}{1015} (\bibinfo{year}{2007}),
  \urlprefix\url{https://link.aps.org/doi/10.1103/RevModPhys.79.1015}.

\bibitem[{\citenamefont{Weickert et~al.}(2010)\citenamefont{Weickert, Brando,
  Steglich, Gegenwart, and Garst}}]{weickert2010universal}
\bibinfo{author}{\bibfnamefont{F.}~\bibnamefont{Weickert}},
  \bibinfo{author}{\bibfnamefont{M.}~\bibnamefont{Brando}},
  \bibinfo{author}{\bibfnamefont{F.}~\bibnamefont{Steglich}},
  \bibinfo{author}{\bibfnamefont{P.}~\bibnamefont{Gegenwart}},
  \bibnamefont{and} \bibinfo{author}{\bibfnamefont{M.}~\bibnamefont{Garst}},
  \bibinfo{journal}{Phys. Rev. B} \textbf{\bibinfo{volume}{81}},
  \bibinfo{pages}{134438} (\bibinfo{year}{2010}),
  \urlprefix\url{https://link.aps.org/doi/10.1103/PhysRevB.81.134438}.

\bibitem[{\citenamefont{Zacharias and Garst}(2013)}]{zacharias2013quantum}
\bibinfo{author}{\bibfnamefont{M.}~\bibnamefont{Zacharias}} \bibnamefont{and}
  \bibinfo{author}{\bibfnamefont{M.}~\bibnamefont{Garst}},
  \bibinfo{journal}{Phys. Rev. B} \textbf{\bibinfo{volume}{87}},
  \bibinfo{pages}{075119} (\bibinfo{year}{2013}),
  \urlprefix\url{https://link.aps.org/doi/10.1103/PhysRevB.87.075119}.

\bibitem[{\citenamefont{Zhao et~al.}(2016)\citenamefont{Zhao, Yelland, Bruin,
  Sheikin, Canfield, Fritsch, Sakai, Mackenzie, and Hicks}}]{zhao2016field}
\bibinfo{author}{\bibfnamefont{L.}~\bibnamefont{Zhao}},
  \bibinfo{author}{\bibfnamefont{E.~A.} \bibnamefont{Yelland}},
  \bibinfo{author}{\bibfnamefont{J.~A.~N.} \bibnamefont{Bruin}},
  \bibinfo{author}{\bibfnamefont{I.}~\bibnamefont{Sheikin}},
  \bibinfo{author}{\bibfnamefont{P.~C.} \bibnamefont{Canfield}},
  \bibinfo{author}{\bibfnamefont{V.}~\bibnamefont{Fritsch}},
  \bibinfo{author}{\bibfnamefont{H.}~\bibnamefont{Sakai}},
  \bibinfo{author}{\bibfnamefont{A.~P.} \bibnamefont{Mackenzie}},
  \bibnamefont{and} \bibinfo{author}{\bibfnamefont{C.~W.} \bibnamefont{Hicks}},
  \bibinfo{journal}{Phys. Rev. B} \textbf{\bibinfo{volume}{93}},
  \bibinfo{pages}{195124} (\bibinfo{year}{2016}),
  \urlprefix\url{https://link.aps.org/doi/10.1103/PhysRevB.93.195124}.

\bibitem[{\citenamefont{Moroni-Klementowicz
  et~al.}(2009)\citenamefont{Moroni-Klementowicz, Brando, Albrecht, Duncan,
  Grosche, Gr\"uner, and Kreiner}}]{moroni2009magnetism}
\bibinfo{author}{\bibfnamefont{D.}~\bibnamefont{Moroni-Klementowicz}},
  \bibinfo{author}{\bibfnamefont{M.}~\bibnamefont{Brando}},
  \bibinfo{author}{\bibfnamefont{C.}~\bibnamefont{Albrecht}},
  \bibinfo{author}{\bibfnamefont{W.~J.} \bibnamefont{Duncan}},
  \bibinfo{author}{\bibfnamefont{F.~M.} \bibnamefont{Grosche}},
  \bibinfo{author}{\bibfnamefont{D.}~\bibnamefont{Gr\"uner}}, \bibnamefont{and}
  \bibinfo{author}{\bibfnamefont{G.}~\bibnamefont{Kreiner}},
  \bibinfo{journal}{Phys. Rev. B} \textbf{\bibinfo{volume}{79}},
  \bibinfo{pages}{224410} (\bibinfo{year}{2009}),
  \urlprefix\url{https://link.aps.org/doi/10.1103/PhysRevB.79.224410}.

\bibitem[{\citenamefont{Kitagawa et~al.}(2005)\citenamefont{Kitagawa, Ishida,
  Perry, Tayama, Sakakibara, and Maeno}}]{kitagawa2005metamagnetic}
\bibinfo{author}{\bibfnamefont{K.}~\bibnamefont{Kitagawa}},
  \bibinfo{author}{\bibfnamefont{K.}~\bibnamefont{Ishida}},
  \bibinfo{author}{\bibfnamefont{R.~S.} \bibnamefont{Perry}},
  \bibinfo{author}{\bibfnamefont{T.}~\bibnamefont{Tayama}},
  \bibinfo{author}{\bibfnamefont{T.}~\bibnamefont{Sakakibara}},
  \bibnamefont{and} \bibinfo{author}{\bibfnamefont{Y.}~\bibnamefont{Maeno}},
  \bibinfo{journal}{Phys. Rev. Lett.} \textbf{\bibinfo{volume}{95}},
  \bibinfo{pages}{127001} (\bibinfo{year}{2005}),
  \urlprefix\url{https://link.aps.org/doi/10.1103/PhysRevLett.95.127001}.

\bibitem[{\citenamefont{Yamase and A.~Katanin}(2007)}]{yamase2007van}
\bibinfo{author}{\bibfnamefont{H.}~\bibnamefont{Yamase}} \bibnamefont{and}
  \bibinfo{author}{\bibfnamefont{A.}~\bibnamefont{A.~Katanin}},
  \bibinfo{journal}{J. Phys. Soc. Jpn.} \textbf{\bibinfo{volume}{76}},
  \bibinfo{pages}{073706} (\bibinfo{year}{2007}).

\bibitem[{\citenamefont{Puetter et~al.}(2010)\citenamefont{Puetter, Rau, and
  Kee}}]{puetter2010microscopic}
\bibinfo{author}{\bibfnamefont{C.~M.} \bibnamefont{Puetter}},
  \bibinfo{author}{\bibfnamefont{J.~G.} \bibnamefont{Rau}}, \bibnamefont{and}
  \bibinfo{author}{\bibfnamefont{H.-Y.} \bibnamefont{Kee}},
  \bibinfo{journal}{Phys. Rev. B} \textbf{\bibinfo{volume}{81}},
  \bibinfo{pages}{081105} (\bibinfo{year}{2010}),
  \urlprefix\url{https://link.aps.org/doi/10.1103/PhysRevB.81.081105}.

\bibitem[{\citenamefont{Lee and Wu}(2009)}]{lee2009theory}
\bibinfo{author}{\bibfnamefont{W.-C.} \bibnamefont{Lee}} \bibnamefont{and}
  \bibinfo{author}{\bibfnamefont{C.}~\bibnamefont{Wu}}, \bibinfo{journal}{Phys.
  Rev. B} \textbf{\bibinfo{volume}{80}}, \bibinfo{pages}{104438}
  (\bibinfo{year}{2009}),
  \urlprefix\url{https://link.aps.org/doi/10.1103/PhysRevB.80.104438}.

\bibitem[{\citenamefont{Raghu et~al.}(2009)\citenamefont{Raghu, Paramekanti,
  Kim, Borzi, Grigera, Mackenzie, and Kivelson}}]{raghu2009microscopic}
\bibinfo{author}{\bibfnamefont{S.}~\bibnamefont{Raghu}},
  \bibinfo{author}{\bibfnamefont{A.}~\bibnamefont{Paramekanti}},
  \bibinfo{author}{\bibfnamefont{E.~A.} \bibnamefont{Kim}},
  \bibinfo{author}{\bibfnamefont{R.~A.} \bibnamefont{Borzi}},
  \bibinfo{author}{\bibfnamefont{S.~A.} \bibnamefont{Grigera}},
  \bibinfo{author}{\bibfnamefont{A.~P.} \bibnamefont{Mackenzie}},
  \bibnamefont{and} \bibinfo{author}{\bibfnamefont{S.~A.}
  \bibnamefont{Kivelson}}, \bibinfo{journal}{Phys. Rev. B}
  \textbf{\bibinfo{volume}{79}}, \bibinfo{pages}{214402}
  (\bibinfo{year}{2009}),
  \urlprefix\url{https://link.aps.org/doi/10.1103/PhysRevB.79.214402}.

\bibitem[{\citenamefont{Mercure}(2008)}]{mercure2008}
\bibinfo{author}{\bibfnamefont{J.-F.} \bibnamefont{Mercure}}, Ph.D. thesis,
  \bibinfo{school}{University of St Andrews} (\bibinfo{year}{2008}).

\end{thebibliography}
\bibliographystyle{apsrev}
\end{document}